\documentclass[acmsmall]{acmart}

\usepackage{booktabs} 
\usepackage{balance,url}
\usepackage{color,soul}
\usepackage{graphicx}
\usepackage{subfig}
\usepackage{multirow}
\usepackage{array}
\usepackage{dsfont}
\usepackage{makecell}
\usepackage{amssymb}

\setcopyright{acmlicensed}
\usepackage{algorithm}
\usepackage[noend]{algpseudocode}
\newcommand*{\Perm}[2]{{}^{#1}\!P_{#2}}%
\newcommand\numberthis{\addtocounter{equation}{1}\tag{\theequation}}

\setcopyright{acmlicensed}
\acmJournal{TDS}
\acmYear{2020} \acmVolume{1} \acmNumber{1} \acmArticle{1} \acmMonth{1} \acmPrice{15.00}\acmDOI{10.1145/3406961}

\begin{document}
\title{WattScale: A Data-driven Approach for Energy Efficiency Analytics of Buildings at Scale}


\author{Srinivasan Iyengar}
\affiliation{Microsoft Research India}
\author{Stephen Lee}
\affiliation{University of Pittsburgh}
\author{David Irwin} 
\affiliation{University of Massachusetts Amherst}
\author{Prashant Shenoy} 
\affiliation{University of Massachusetts Amherst}
\author{Benjamin Weil}
\affiliation{University of Massachusetts Amherst}
\email{}

\begin{abstract}

Buildings consume over 40\% of the total energy in modern societies, and improving their energy efficiency can significantly reduce our energy footprint. In this paper, we present \texttt{WattScale}, a data-driven approach to identify the least energy-efficient buildings from a large population of buildings in a city or a region. Unlike previous methods such as least-squares that use point estimates, \texttt{WattScale} uses Bayesian inference to capture the stochasticity in the daily energy usage by estimating the distribution of parameters that affect a building. Further, it compares them with similar homes in a given population. \texttt{WattScale} also incorporates a fault detection algorithm to identify the underlying causes of energy inefficiency. We validate our approach using ground truth data from different geographical locations, which showcases its applicability in various settings. \texttt{WattScale} has two execution modes --- (i) individual, and (ii) region-based, which we highlight using two case studies. For the individual execution mode, we present results from a city containing >10,000 buildings and show that more than half of the buildings are inefficient in one way or another indicating a significant potential from energy improvement measures. Additionally, we provide probable cause of inefficiency and find that 41\%, 23.73\%, and 0.51\% homes have poor building envelope, heating, and cooling system faults, respectively. For the region-based execution mode, we show that \texttt{WattScale} can be extended to millions of homes in the US due to the recent availability of representative energy datasets.   

\end{abstract}

%
%




\maketitle

\section{Introduction}
\label{sec:introduction}

Buildings constitute around 40\% of total energy and 70\% of the overall electricity usage in the United States~\cite{doereport}.
Consequently, building energy-efficiency has emerged as a significant area of research in smart grids. A typical city comprises a large number of buildings of different sizes and age. In general, the building stock in many North American and European cities tend to be old---while some are recently constructed, the majority were built decades ago. Moreover, it is not uncommon for buildings to be over a hundred years old~\cite{doereport}. 
Technological advances in building construction have yielded better-insulated envelopes as well as more energy-efficient air-conditioning, heating furnaces, and appliances, which can reduce the total energy consumption of a building. While newer buildings, as well as older ones that have undergone renovations, have adopted such efficiency measures, most are yet to benefit from such efficiency improvements.
Since roughly half of a building's energy usage results from heating and cooling, opportunities abound for making efficiency improvements 
in cities around the world.

Since a city may consist of thousands of buildings, an essential first step for implementing energy-efficiency measures is to identify those that are the least efficient and thus have the greatest need for energy-efficiency improvements. Interestingly, naive approaches such as using the age of the building or its total energy bill to identify inefficient buildings do not work well. While older buildings are usually less efficient than newer ones, the correlation is shown to be weak~\cite{chung2006benchmarking}. Thus, \emph{age alone is not an accurate indicator of efficiency}, since older buildings may have undergone renovations and energy improvements. Similarly, the total energy usage is not directly correlated to energy inefficiency. First, larger buildings will consume more energy than smaller ones. Even normalizing for size, greater energy usage does not necessarily point to inefficiencies. 
For example, a single-family home will have a higher energy demand (possibly due to the in-house washer, dryer, and water heater) compared to an identically sized apartment home. 
Thus, finding truly inefficient buildings requires more sophisticated methods. 
	
In this paper, we present a data-driven approach for determining the least efficient residential buildings from a large population of buildings within a city or a region using energy data in association with other external public data sources. 
Such buildings can then become candidates for energy efficiency measures including 
targeted energy incentives for improvements or upgrades. 
So far, lack of granular city-wide datasets prevented large-scale energy efficiency analysis of buildings. 
However, with increasing smart meter installations across a utilities' customer base, energy usage information of buildings is readily available. By 2016, the US had more than 70 million installed smart meters (>700M worldwide)~\cite{engie}. 
Also, real estate information describing a building's age, size, and other characteristic are public records in many countries.
Further, weather conditions can be accessed through REST APIs. Reliance on such readily available datasets make our approach broadly applicable. 


Given these datasets, 
our approach assumes it is possible to model a building's total energy usage as a sum of \emph{weather-dependent} and \emph{weather-independent} energy components.  
The weather-dependent component captures the heating and cooling energy usage, which is typically a function of the external temperature, while the weather-independent component captures the energy use from all other activities.   
Using this approach, we can then extract the parameter distributions that govern these energy components and identify causes of energy inefficiency by 
comparing them to those of other homes in a given population.
For example, a model's parameter that is more sensitive to external temperature is indicative of inefficient heating or cooling.  
We also develop algorithms that use these comparisons to determine the probable causes of energy inefficiency.

While building energy models have been extensively studied in the energy science research for many decades \cite{
thom1954rational, fels1986prism, allen1976modified}, and practitioners such as energy auditors routinely use them to analyze a building's energy performance, there are important differences between current approaches and our technique. 
First, current models employ several important parameters that are often chosen manually, based on rules of thumb~\cite{jacobs2002state}. 
However, using manually chosen parameters may lead to incorrect analysis~\cite{brown2014climate}.
On the other hand, our technique determines a custom parameter distribution of the building model, and we experimentally show 
its efficacy over manual approaches. 
Second, the current energy models are based on least-squares regression analysis 
that provides point estimates. In contrast, 
our approach provides Bayesian estimates to determine building parameter distribution that captures the stochasticity in energy use. 
Third, current approaches need manual intervention to varying degrees to interpret model parameters and determine likely efficiency issues. Clearly, this does not scale to thousands of buildings across a city. 
Our technique automates this process by comparing model parameters with similar homes from the population and makes it feasible to perform large-scale analysis. Thus, we go beyond determining which buildings are inefficient by also designing algorithms that determine its probable causes.


In this paper, we introduce \texttt{WattScale}, a data-driven approach to determine the most inefficient buildings 
present in a city or a region.
Our contributions are as follows:
\\
{\bf Bayesian Energy Modeling Approach.} 
\texttt{WattScale} improves over prior work that provides point estimates by using a Bayesian inference to capture the building model parameter distributions that govern the energy usage of a building.
These distributions are compared using \textit{second-order stochastic dominance} to create a partial order among building parameters.
Further, we propose a fault analysis algorithm that utilizes these partial orders to report probable causes of inefficiency. 
\\
{\bf Open-source tool with Dual Execution Modes.} We implement \texttt{WattScale} approach as an \emph{open source} tool that enables determining inefficient buildings at scale. Our tool offers two execution modes --- (i) individual, and (ii) region-based. In the individual execution mode, we flag inefficient homes by comparing their building model parameter distributions with other similar homes in a city. Whereas in the region-based execution mode, we compare the building model parameter distributions of the candidate home with those learned for the entire population of similar homes in a given region with similar weather conditions. 
\\
{\bf Model Validation and Analysis.} We evaluate \texttt{WattScale} using energy data from three different cities in geographically diverse regions of the US. 
In particular, we show that 
our approach can disaggregate  
the buildings' energy usage into different components with high accuracy and tighter bounds on the model parameters 
 --- an improvement over the two popular baselines. 
Further, our approach identifies buildings that have possible energy inefficiencies. 
In comparison to manual audit reports, our approach correctly identified faults in nearly 95\% of the cases. 
\\
{\bf Real-world case study analysis and wide applicability.} 
We examine our approach using two different case studies showcasing the efficacy of the two execution modes of \texttt{WattScale}. In the first case study, we used the individual execution mode as we had energy usage from smart meters deployed in 10,107 residential buildings in a city through a local utility. 
\texttt{WattScale} reported more than half  of the buildings in our dataset as inefficient, which indicates a significant scope for making energy improvements in 
several cities. 
Further, our results indicate poor building envelope as a major cause of inefficiency, which accounts for around 41\% of all homes. 
Heating and cooling system faults comprises 23.73\%, and 0.51\% of all homes respectively. In another case study, we used region-based execution mode on a smaller dataset of residential buildings from the city of Boulder. Here, we showed that region-based mode can help detect faults in millions of residential buildings in the US and around the world, if a representative energy dataset if available.   
Thus, using the region-based mode, the individual homeowners can proactively learn about the energy efficiency of their homes without the intervention from their local utility. 

\section{Background}
\label{sec:background}
In this section, we present background on energy efficiency in buildings and 
techniques 
used to model a building's energy usage. 

\subsection{Energy Efficiency in Buildings}
\begin{figure}[t]
\centering
\begin{tabular}{cc}
\includegraphics[width=1.6in]{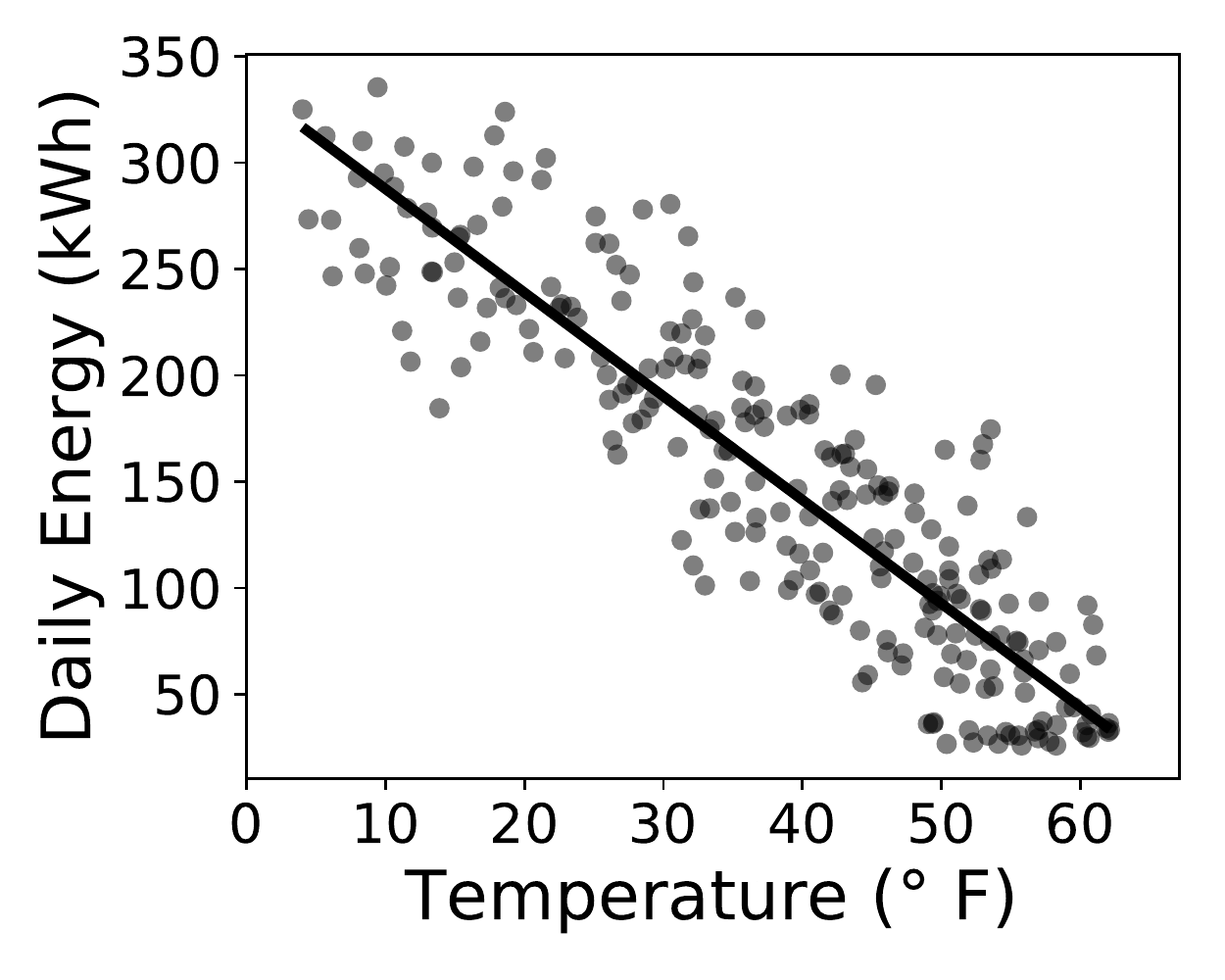} &
\includegraphics[width=1.6in]{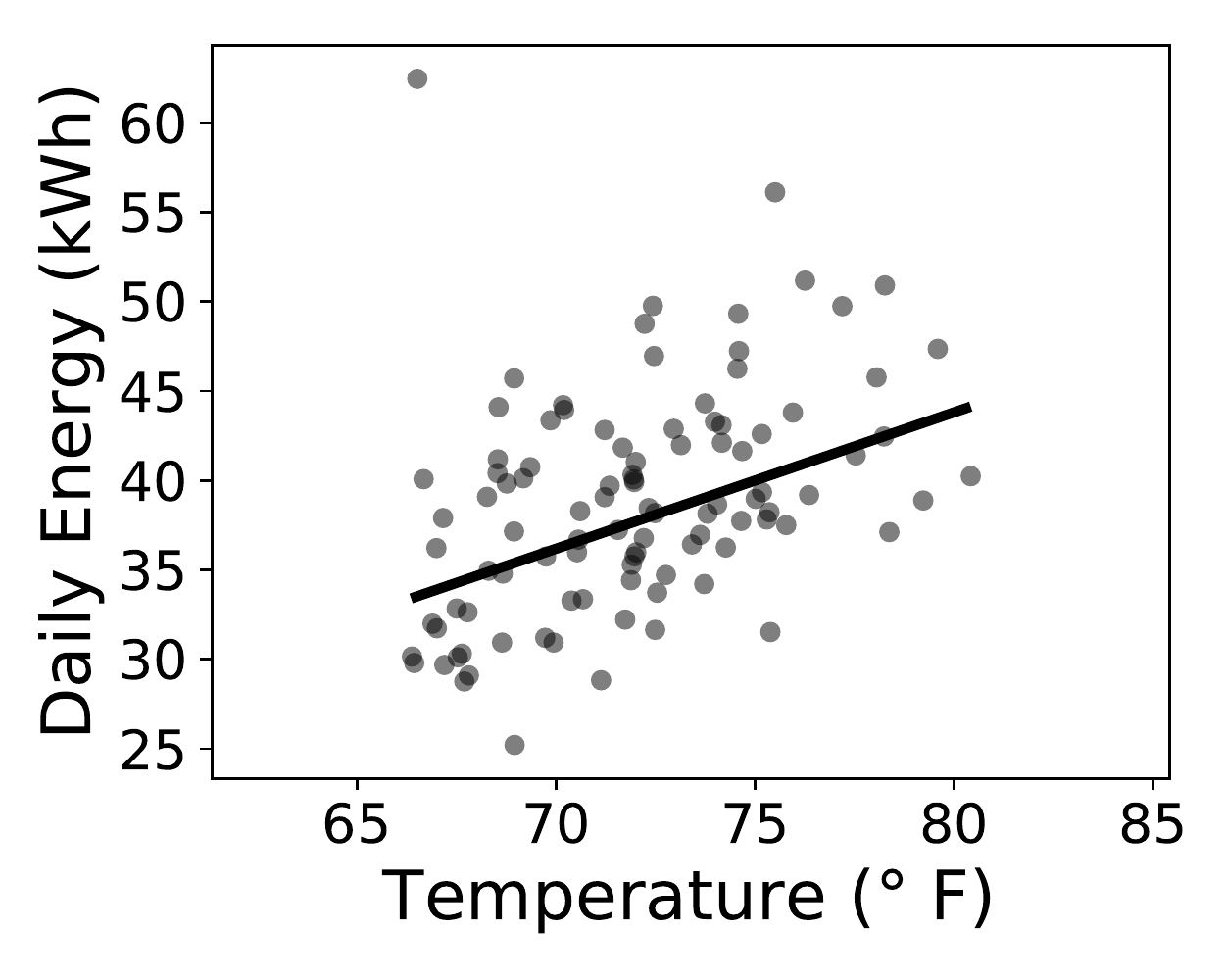}  \\
(a) Winter Months & (b) Summer Months\\
\end{tabular}
\caption{Linear relationship between energy consumption and ambient temperature for a Single Family home from the New England region of the US (energy audit year: 2015)}
\label{fig:example-regression}
\end{figure}

Energy usage in residential buildings  has different sources such as heating and cooling, lighting, household appliances etc. There can be many causes of inefficiencies in each of these components, such as the use of inefficient incandescent lighting and the use of inefficient (e.g., non-energy star) appliances. 
Studies have shown that  heating and cooling is the dominant portion of a building's energy usage, comprising over half of the total usage \cite{eia-statistics,doereport}, and it follows that the most significant cause of inefficiency lies in problems
with heating and cooling. Two factors determine heating and cooling efficiency of a building: (1) the insulation of the building's external walls and roof ("building envelope") and their ability to minimize thermal leakage, and (2) the efficiency of the heating and cooling equipment. Recent technology improvements have seen advancements on both fronts. New buildings are constructed using modern methods and better construction materials that yield a building envelope that minimizes air leaks and thermal loss through better-insulated walls and roofs and high-efficiency windows and doors. Similarly, new high-efficiency heating and AC equipment are typically 20-30\% more efficient than equipment typically installed in the late 1990s and early 2000s. 

Unfortunately, older residential buildings and even ones built two decades ago do not incorporate such energy efficient features. Further, the building envelope can deteriorate over time due to age and weather and so can mechanical HVAC equipment. 
Consequently, an analysis of a building's heat and cooling energy use can point to the leading causes of a building's energy inefficiency.

\subsection{Inferring a Building Energy Model}
One approach to modeling a building's heating and cooling usage is to model its dependence on weather~\cite{zhao2012review}. For example, a building's heating and cooling usage can be modeled as a linear function of external temperature. To intuitively understand why, consider cooling energy usage during the summer. The higher the outside temperature on hot summer days, the higher the AC energy usage. 
Since the difference between outside and inside temperatures is large, there is more thermal gain, which requires longer duration of cooling to maintain a set indoor temperature. Thus, there is a linear relationship between heating/cooling energy use and outside temperature (see Figure~\ref{fig:example-regression}(a) and (b)). 
Given the linear dependence, linear models are commonly used within the energy science research~\cite{kissock2002development,fels1986prism}, 
to capture the relationship between energy use and outside temperature.  
However, most of the prior approaches do not consider uncertainties that are associated with indicators of building performance. 
Primarily, these models do not capture the stochastic variations in heating and cooling as well
as the weather-independent energy usage resulting from day to day variations in human activities inside a home. 
As seen in Figure~\ref{fig:example-regression}, such energy variations exist 
and our approach uses Bayesian inference to determine the distributions of the building parameter 
that models these uncertainties in energy use.

\subsection{Problem Formulation}

Consider a large population of buildings in a city. 
We assume that a trace of the total daily energy usage is available for each building. 
We also assume building characteristics, such as age, size, and type (Single Family, Apartment etc.) for each building along with the daily outdoor temperature data are available. 

Let $B$ be the set of all residential buildings containing information on building characteristics in a city. 
Further, $b_{i} \in B$ 
denotes the $i^{th}$ residential building defined by a tuple $\langle E^{total}_{i, [1...D]}, Age_{i}, Size_{i}, Type_{i} \rangle$. Here, $E^{total}_{i, [1...D]}$ is the energy usage recorded by smart meters for a period of $D$ days. Moreover, $T_{[1...D]}$ is the external ambient temperature for the city during the $D$ days. Thus, given $b_{i}  \in B$ and $T_{d} \forall d \in D$, our problem is to determine $(a_{1}, ..., a_{m})_{i} \in \{False,True\}^{m}$, where $a_{1}, ..., a_{m}$ are the $m$ possible faults associated with the residential buildings. 

\section{WattScale: Our Approach }
\label{sec:design}

\begin{figure}[t]
\centering
\includegraphics[width=3in]{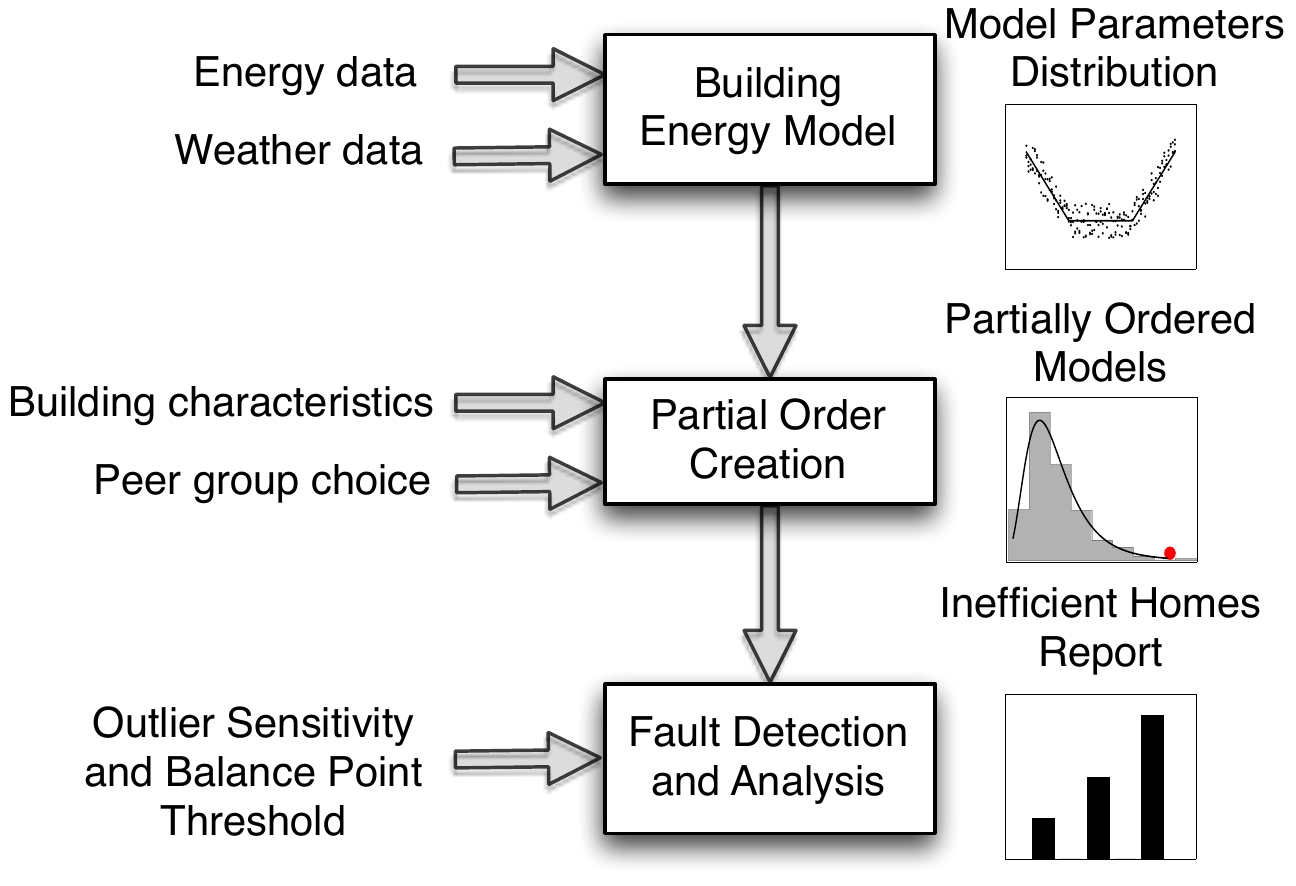}
\caption{Overview of \texttt{WattScale} approach.}
\label{fig:block_diagram}
\end{figure}

In this section, we describe the details of our data-driven approach. \texttt{WattScale}'s approach is 
depicted in Figure~\ref{fig:block_diagram} and it 
involves
three key steps: 
(i) Learn a \emph{building energy model} for a home or a region from energy usage data,
 (ii) Create a \emph{partial order} of buildings using its parameter distribution from the building model, and finally 
 (iii) Detect \emph{building faults} causing energy inefficiency for a home. Below, we discuss each step in detail.

\subsection{Building Energy Model}
\begin{figure}[t]
\centering
\includegraphics[width=3in]{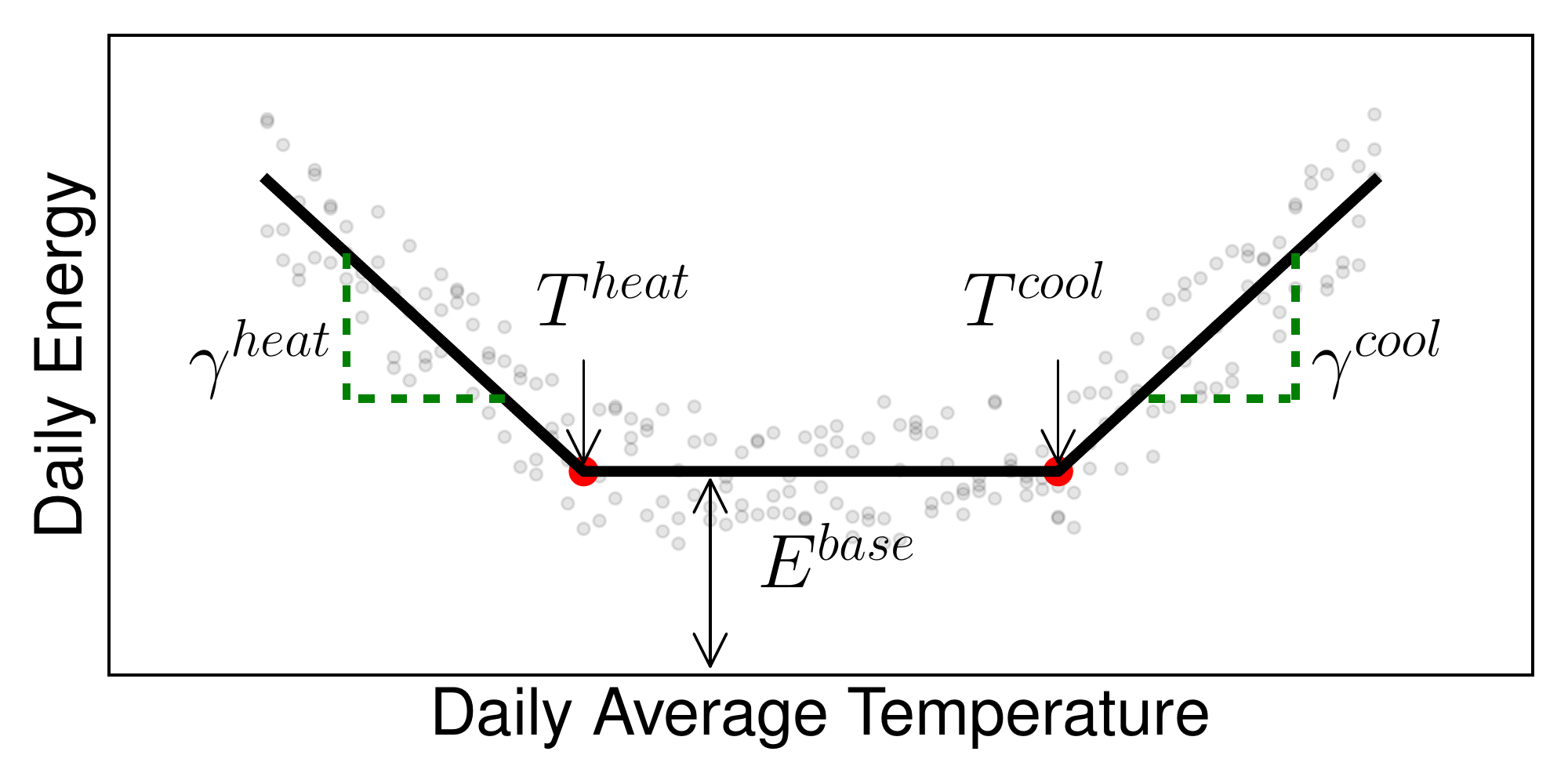}
\caption{An illustrative figure showing energy usage versus ambient outdoor temperature.}
\label{fig:degreedays}
\end{figure}

We first provide the intuition behind our approach. 
Heating and cooling costs for a building can be understood using elementary thermodynamics. Typically, in colder months, the outside ambient temperature is colder than the inside building temperature, resulting in a net thermal loss where the inside heat flows outside through the building envelope, causing the inside temperature to drop. In warmer months, the opposite is true. The building experiences a net heat gain where the heat flows inside, causing the building temperature to rise. 

It follows that every home has a specific temperature $T_{b}$, where there is neither thermal loss nor thermal gain i.e. the thermodynamic equilibrium. When the outside temperature is above $T_{b}$, there is a need for AC to cool the home. Conversely, when the temperature is below $T_{b}$, there is a need for a heater to heat the home. This temperature $T_{b}$ is called the balance point temperature of the building~\footnote{Note that this balance point temperature is not the indoor thermostat setpoint temperature of the building. It is merely a thermodynamic construct where the heat transfer between the building and the outdoor environment is zero.}. The rate of thermal loss or thermal gain depends on the degree of insulation, airtightness of the building envelope and surface area exposed to outside elements. Better the insulation and airtightness, smaller the rate of loss or gain for a given temperature differential relative to $T_{b}$. The difference between the outside temperature and the balance point temperature $T_{b}$ is also referred as the \textit{degree-days} --- an indication of how many degrees warmer or colder is the outside weather relative to the building's balance point.

Based on this intuition, we now describe our building energy model. Any energy load in a building can be classified as weather independent and dependent. A weather independent load is one where the energy consumed by  the device is uncorrelated to the outside temperature --- consumption from loads such as lighting, electronic devices, and household appliances depend on human activity rather than outside weather. Heating and cooling equipment constitute weather dependent loads, as their consumption linearly dependent on the outside temperature relative to the balance point. 

If we assume that weather independent loads are distributed around a constant value (also called the base load); then the total energy consumed is the sum of the base load and the weather dependent loads (heating and cooling loads) and defined as: 
\begin{equation}
E^{total}_{d} = E^{heat}_{d} + E^{cool}_{d}  + E^{base}  \quad \quad \forall d \in D
\label{eqn:energymodel}
\end{equation}
where $E^{total}_{d}$ denotes the total energy used by a building on day $d \in D$. 
$E^{heat}_{d}$ and $E^{cool}_{d}$ denote the energy used for heating and cooling, respectively, on day $d$, while $E^{base}$ denotes the energy usage of base load appliances. 
Thus, given a series of observations of the total energy usage and the outside ambient temperature, it is possible to fit a regression and learn the fixed weather independent component (base load) and the temperature dependent component (heating and cooling). This forms the basis for inferring our weather-aware building energy model. 

Figure~\ref{fig:degreedays} illustrates the relationship between 
outdoor temperature and the energy consumption of a building. The individual data points represent the daily energy usage (along the Y-axis) for a given average 
outdoor temperature (along the X-axis) of a building. The figure shows that the building has two balance point temperatures --- a heating balance point temperature $T^{heat}$, below which heating units are turned on, and a cooling balance point temperature $T^{cool}$, above which air-conditioning is turned on. Further, the figure also shows a piecewise linear fit over the daily energy usage. When the outdoor temperature is between the two balance points, the building consumes energy that is distributed around a constant value defined as the \emph{base load} $E^{base}$ energy consumption. The weather dependent components, i.e. the heating $E^{heat}$ and cooling $E^{cool}$ energy consumption, are a function of ambient outdoor temperature $T_{d}$ and are defined as:
\begin{align}
E^{heat}_{d} &= \gamma^{heat} (T^{heat} - T_{d})^+ \quad \forall d \in D
\label{eqn:heat}\\
E^{cool}_{d} &= \gamma^{cool} (T_{d} - T^{cool})^+ \quad \forall d \in D
\label{eqn:cool}
\end{align} 
where $\gamma^{heat}$ and $\gamma^{cool}$ are the heating and the cooling slope in the above linear equations and represent a positive constant factor indicating the sensitivity of the  building to temperature changes; and
$()^+$ indicates the value is zero if negative and ensures either energy from heating or cooling is considered. 
Using (\ref{eqn:heat}) and (\ref{eqn:cool}), energy model in~(\ref{eqn:energymodel}) can be represented as a piecewise linear model:

\begin{align*}
E^{total}_{d} = & E^{base} + \gamma^{heat} (T^{heat} - T_{d})^+  +  \gamma^{cool} (T_{d} - T^{cool})^+    \forall d \in D  \numberthis
\label{eqn:model}
\end{align*} 
The model in (\ref{eqn:model}) is known as the \emph{degree-day} model~\cite{kissock2002development} 
and forms our base energy model for estimating the building parameters. A more in-depth explanation is presented in ASHRAE guideline 14 referring to the five-parameter change point model~\cite{guideline2014guideline}. Note that the above model will work when data for at least a year is available. However, a truncated version of the model can be employed when only heating (cooling) data is available for winter (summer) months.


\subsubsection{Bayesian Inference Parameter Estimation of a Building}
\label{sec:paramestimation}
While methods like Maximum Likelihood Estimation (MLE) or Maximum a posteriori estimation (MAP) can be used for determining the building parameters, 
they provide point estimates that can hide relevant information (such as not capturing the uncertainties in human energy usage). 
To capture human variations, we require probability density function of the parameters. 
Thus, we use Bayesian inference approach, which provides the posterior distribution of parameters.  


We model (\ref{eqn:model}) using a bayesian approach and assume the error process to be normally distributed ($\mathcal{N}$(0, $\sigma^2$)). Thus, the daily energy consumption $E_d^{total}$ is normally distributed with parameters mean ($\mu$) and variance ($\sigma^2$), where $\mu$ is equal to the right hand side of (\ref{eqn:model}).  Note that energy consumption $E_d^{total}$ is known and so is the independent variable i.e. ambient temperature $T_d$. 
However, the building parameters ($\gamma^{heat}$, $\gamma^{cool}$, $T^{heat}$, $T^{cool}$, and $E^{base}$) are unknown.
 Using Bayesian inference, we can then compute a \emph{posterior} distribution 
 for each of these parameters that best explains the \emph{evidence} (i.e. the known values for $E^{total}_{d}$ and $T_{d} \forall d \in D$) from initially assuming a \emph{prior} distribution. 

To determine the posterior distribution of the individual parameters, we use the
Markov chain Monte Carlo (MCMC) method that generates samples from the posterior distribution by forming a reversible Markov-chain with the same equilibrium distribution. 
We introduce a prior distribution that represents
 the initial belief regarding the building parameters. For example, the two balance point temperatures will be between a wide range of 32\textdegree F and 100\textdegree F. This belief can be represented using a uniform prior with the said range. Similarly, the baseload, heating slope and cooling slope can be drawn from a weakly informative gaussian prior having non-zero values. This is because baseload, a unit of energy, cannot be negative. Similarly, slope values must be positive as they represent increase in energy per unit temperature. The parameters of the gaussian priors are scaled to our setting and selected based on the recommendations provided by Gelman et al.~\cite{gelman2006prior}. To simplify our building model, we assume that the parameters are independent, i.e., the heating, cooling and the baseload parameters do not affect one another.

Several MCMC methods leverage different strategies to lead from these priors towards the target posterior distribution. We employed No-U-turn sampler,
a sophisticated MCMC method, which has shown to converge quickly towards the target distribution. Thus, after an initial \emph{burn in} samples, we can draw samples approximating the true posterior distribution. From these post-burn-in samples, a posterior distribution for the individual building parameters can be formed. Our complete Bayesian model is defined in Table~\ref{tbl:model}. 

\begin{table}[]
\centering
\begin{tabular}{l} \hline
\textbf{Prior }             \\  
$E^{base} \sim \mathcal{N}(20, 20)$,  $\gamma^{heat} \sim \mathcal{N}(0, 4)$,  $\gamma^{cool} \sim \mathcal{N}(0, 4)$ \\ 
$T^{heat} \sim\mathcal{U}(32, 100)$,  $T^{cool} \sim\mathcal{U}(32, 100)$, $\sigma \sim Cauchy(0, 5)$ \\ 
\textbf{Regression Equation}\\
$\mu_{d}$ = $E^{base}$ + $\gamma^{heat} (T^{heat} - T_{d})^+$ + $\gamma^{cool} (T_{d} - T^{cool})^+$ $ \forall d \in D$ \\ 
\textbf{Model Likelihood}\\
$E^{total}_{d} \sim \mathcal{N}(\mu_{d}, \sigma^{2})$\\ 
\textbf{Parameter Bounds} \\
$E^{base},  \gamma^{heat}, \gamma^{cool} >= 0 \quad \text{and} \quad T^{heat} <=T^{cool} $\\ \hline
\end{tabular}
\vspace{0.1in}
\caption{Bayesian formulation of our building energy model.}
\label{tbl:model}
\end{table}

Since buildings are of different sizes, simply comparing  the parameters in absolute terms is not meaningful. To enable such comparison, we initially normalize the energy usage by building size before the Bayesian inference. Hence, in our case, $E^{base}$ represents base load energy use per unit area. Similarly, heating slope $\gamma^{heat}$ and cooling slope $\gamma^{cool}$ gives change in energy per degree temperature per unit area. The balance point parameters ($T^{heat}$ and $T^{cool}$) are not normalized as they are unaffected by the size of the house. We construct a cumulative distribution ($F_{\gamma^{heat}}$, $F_{\gamma^{cool}}$, $F_{E^{base}}$) for each of the building model parameter ($\gamma^{heat}$, $\gamma^{cool}$, $E^{base}$) from their respective density functions (posterior) obtained after the inference. For the balance point parameters ($T^{heat}$ and $T^{cool}$), we only use its mean values as they tend to remain fixed for a given building irrespective of human variation. 

\subsubsection{Building Parameter Estimation of a Region}
The building energy model in (\ref{eqn:model}) can also be used to the estimate the building parameters for a region. Estimating the distribution of building parameters of a region can allow efficient comparison of a building to a general population. Here, we describe how we can create the building energy model for a given region. 
Since the above model uses daily energy usage for each home, estimating the parameter distribution for an entire population may be inefficient and time-consuming. 
Further, such fine-grained daily energy usage for all homes in a region may not be available. 
Instead, we use the annual consumption information to estimate the population's building parameter~\cite{bpd}. To do so, we modify our energy model as follows. 
Similar to (\ref{eqn:heat}) and (\ref{eqn:cool}), the weather component of a building can be defined as.
\begin{align}
E^{heat}_{h} &= \gamma^{heat}_h \sum_{d}^{D} (T^{heat}_h - T_{d})^+ \quad \forall h \in H
\label{eqn:heatall}\\
E^{cool}_{h} &= \gamma^{cool}_h \sum_{d}^{D}  (T_{d} - T^{cool}_h)^+ \quad \forall h \in H
\label{eqn:coolall}
\end{align} 
where $H$ is a set of homes in a region and $E^{heat}_{h}$ and $E^{cool}_{h}$ are the annual heating and cooling consumption for home $h$. 
Further, the energy model of a home $h$ can be represented as. 

\begin{align*}
E^{total}_{h} = & E^{base} \cdot |D|+ \gamma^{heat}_h \sum_{d}^{D}  (T^{heat}_h - T_{d})^+  +  \gamma^{cool}_h \sum_{d}^{D}  (T_{d} - T^{cool}_h)^+    \quad \forall h \in H  \numberthis
\label{eqn:modelall}
\end{align*} 
where $E^{total}_{h}$ is the total annual energy consumption of home $h$. This forms the base energy model for estimating the building parameters of a home in a region. 

In the equations~\ref{eqn:heatall}, \ref{eqn:coolall}, \ref{eqn:modelall}, there are five unknowns per home associated with  the five parameters ($\gamma^{heat}_{h}$, $\gamma^{cool}_{h}$, $E^{base}_{h}$, $T^{heat}_{h}$, and $T^{cool}_{h}$). By assuming $T^{heat}_{h} = T^{cool}_{h} =$ 65~\textdegree F, we can estimate the other parameters by solving these equations using the known annual energy consumption values available per home --- i.e., $E^{heat}_{h}$, $E^{cool}_{h}$ and $E^{total}_{h}$. Next, we construct a cumulative distribution ($\hat{F}_{\gamma^{heat}}, \hat{F}_{\gamma^{cool}}, \hat{F}_{E^{base}}$) for each of the building model parameter ($\gamma^{heat}$, $\gamma^{cool}$, $E^{base}$) of the input region using the \textit{Kernel Density Estimation}~\footnote{This is also called Parzen-Rosenblatt window method.}
, a popular nonparametric approach to estimate a random variable. Later, we will show how we use this parameter distribution of a region to identify an inefficient home.

\subsection{Partial Order Creation} 
\label{sec:outlierdetect}
Rather than relying on rule-of-thumb measures to interpret model parameters that change with geography and many other building characteristics, we propose comparing them with those of similar homes from a given population. 
Given the above model, we create a partial order of buildings as follows.
We first create \emph{peer groups} using the building's physical attributes (e.g., age of the building, building type etc.).
Next, within each peer group 
we create a  \emph{partial order} of the buildings for each building parameter distribution. 
Below, we describe each step in detail.

\subsubsection{Peer groups creation}

A naive approach of comparing model parameters of any two homes has several shortcomings. First, building parameters may vary based on the building type. 
As an example, consider the energy use of a studio apartment and a three-bedroom apartment. Both building type have completely different energy needs in term of cooling/heating loads, and the rate of heat gain (or loss) would be different.
Hence, a building model's heating/cooling parameter from two different building type would be different, and thus, should not be compared in the same cohort.
Second, even for the same building type, the model parameters from two buildings built in a different year, may belong to two different families of distributions, and thus may be an unfair comparison.  As an example, assume two houses are equal in all aspects (building characteristics, occupancy patterns, etc.) except year built. Due to advances in building technology and energy efficiency standards, a newer home will have building envelopes made using more energy efficient material than a comparatively older home. While the newer home may be energy efficient compared to older homes, the newer home may still be energy inefficient compared to cohort of homes built around the same year. Thus, it would be unreasonable to compare the building model parameters from homes with a sizeable age difference as outlier detection techniques will always mark older homes as inefficient. To overcome some of these limitations, \texttt{WattScale} allows the creation of peer groups to allow comparison within a cohort to determine inefficient homes.

To enable a meaningful comparison, we compare the building model parameters only within their cohort. 
We use three building attributes for peer group creation namely: (i) property class (e.g., single family, apartment, etc.), (ii) built area (e.g., 2000 to 300 sq.ft.), and (iii) year built (e.g.1945 to 1965). For instance, buildings constructed in different years adhere to different energy regulations and standards, and thus, it is not meaningful to compare them. 
Similarly, building types and age group have different characteristics and it would be unreasonable to compare them. Hence, our approach allows the creation of peer groups to enable comparison within a cohort to determine inefficient homes.




\subsubsection{Stochastic order of building parameters}
Since the building model parameters are probabilistic distributions, we cannot simply compare these uncertain quantities and create a \emph{total ordering}. 
Statistics, such as mean, median or mode, provide a single number to capture the behavior of the whole distribution. While these \textit{point estimates} can be used to compare two distributions, they typically hide useful information regarding their shape and may not account for any heavy-tailed nature that is present in a building parameter distribution. 
Hence, we use \emph{second order stochastic dominance}, a well-known concept in decision theory for comparing two distributions~\cite{levy2015stochastic}, to create a partial order of the building parameters within a peer group.  

The main idea behind determining \emph{second order stochastic dominance}
is that for a given building model parameter $p$, if distribution  $F_p$ dominates $G_p$ i.e., $F_p \succeq_{2} G_p$, then the area enclosed between $F_p$ and $G_p$ distribution should be non-negative up to every point in $x$: 
\begin{equation}
\int_{a}^{x}(G_p(t) - F_p(t)) dt \ge 0  \quad \quad \forall x \in [a,b]
\end{equation}
\begin{figure}[t]
\includegraphics[width=3.2in]{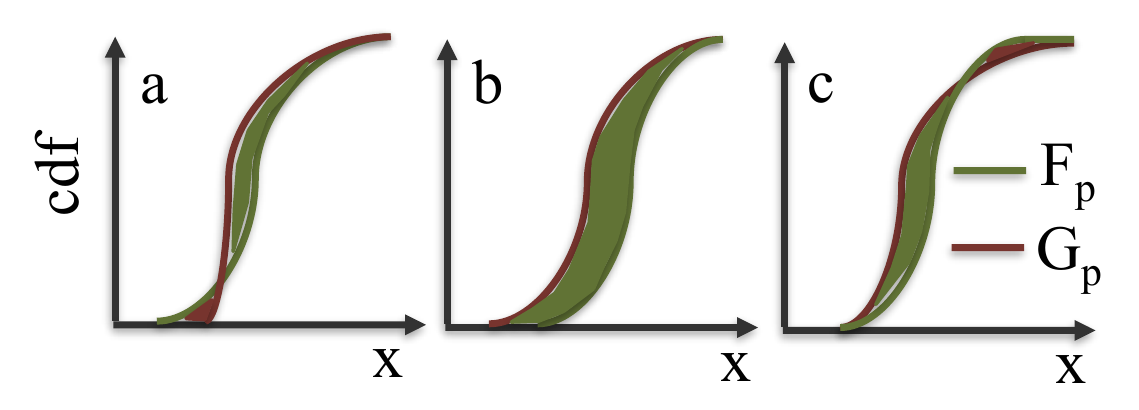} 
\caption{Stochastic ordering of two distributions $F_p$ and $G_p$. (a) $F_p$ does not dominate $G_p$. In (b) and (c) $F_p$ dominates $G_p$.}
\label{fig:stochastic}
\end{figure}
Figure~\ref{fig:stochastic} depicts stochastic ordering of two distribution $F_p$ and $G_p$ where; (i) $F_p$ does not dominate $G_p$ i.e. $F_p \nsucceq_{2} G_p$ and (ii) $F_p$ dominates $G_p$ i.e., $F_p \succeq_{2} G_p$. The area shaded in green shows the region where $F_p$ dominates $G_p$, and the red region shows $G_p$ dominates $F_p$. In  Figure~\ref{fig:stochastic}(a),
we observe that  $F_p \nsucceq_{2} G_p$, since there are no green area greater or equal to the left of the red area. In contrast, Figure~\ref{fig:stochastic}(b) and (c) shows $F_p$ dominates $G_p$ because for every red area, there exists a larger green area located to its left. 

To intuitively understand the implications of stochastic dominance in our scenario, let us consider two distributions $F_p$ and $G_p$ of a building parameter $p$ from two separate buildings $A$ and $B$ respectively. As noted earlier, building parameters influences energy usage, such that higher parameter values implies higher energy usage, and vice-versa. Let us    assume that building $A$'s normalized energy usage is greater than building $G$'s normalized energy usage, such that distribution $F_p$ dominates $G_p$ i.e.,  $F_p \succeq G_p$. Clearly, the building parameter distribution $F_p$ for building $A$ will lie on the right-side of distribution $G_p$ as $A$ has higher energy usage. In fact, since $F_p \succeq G_p$, by definition, the distribution $F_p$ will be on the right of $G_p$ for a majority of the region. 
However, homes may have similar building parameter distribution, i.e the distribution has similar shape and tendency. In such cases, it is possible that neither home will dominate the other. 
Stochastic dominance thus enables interpretation of the building parameter distribution with respect to one another, with higher energy usage buildings having a tendency to lie on the right side of the population. This allows separation of homes with dominant distributions from non-dominant ones. 

\subsubsection{Dual execution modes}
Our \texttt{WattScale} approach can be used in two execution modes --- (i) individual and (ii) region-based. In the individual approach, we run a pair-wise comparison of all buildings within a cohort for each building model parameter $p$. This gives us the partial order for all pairs and parameters, which we use to detect inefficient homes. 
In the region-based approach, we compare the building model parameter to that of the region's parameter distribution. 

Using the stochastic order criteria, it is simple to compare two distributions and identify the dominant distribution. But, there may be cases where there aren't sufficient buildings in a region to create a building parameter distribution of a region. This is because energy consumption data for a given cohort may be sparse. A small city or a region may not have enough buildings to create a parameter distribution for that region and cohort. In order to handle such cases, one approach is to use candidate buildings from nearby regions to create a region-wide parameter distribution for comparison. 

In our approach, we use an R-tree based data structure to access buildings within a region. R-tree data structures can provide efficient access to spatial objects, especially geographical coordinates. The key idea is that the data structure groups nearby homes and represent them with their minimum bounding rectangle. At the leaf level, each rectangle can be represented as a tree, and subsequent aggregation at higher levels, combine nearby objects, providing a coarse approximation of the data. 
Thus, it provides fast and efficient access to a group of homes for any region within the bounding rectangle, and the search area can be increased as needed. In our approach, the search space is increased if we do not find sufficient homes to create a building parameter distribution of a region that meets the specified filter criteria. 
For instance, R-tree can be used to retrieve all homes within a region that were built within a specific year and are of a particular property type (e.g., single family homes). If there are not enough homes that meet the criteria, we include buildings from nearby regions such that the climate conditions of these areas are similar. 
After sufficient buildings are found, we use these buildings to create the parameter distribution of the region for that peer group. 
\subsection{Fault Detection and Analysis}
We first discuss the causes of inefficiencies associated with the different model parameters. Later, we present our algorithm that identifies inefficient homes and its potential cause. 

\begin{table}[]
\centering \small
\begin{tabular}{|l|l|}
\hline
\multicolumn{1}{|c|}{\textbf{Indicator Characteristics}} & \multicolumn{1}{c|}{\textbf{Probable Building Faults}}                                                          \\ \hline
High Heating Slope                       & \begin{tabular}[c]{@{}l@{}}Inefficient Heater,  Poor Building Envelope\end{tabular} \\ \hline
High Cooling Slope                       & \begin{tabular}[c]{@{}l@{}}Inefficient AC,   Poor Building Envelope\end{tabular}     \\ \hline
High Heating Balance Point               & \begin{tabular}[c]{@{}l@{}}High Set point, Poor Building Envelope\end{tabular}      \\ \hline
Low Cooling Balance Point                & \begin{tabular}[c]{@{}l@{}}Low Set point,  Poor Building Envelope\end{tabular}       \\ \hline
High Base load                           & Inefficient Appliances                                                                        \\ \hline
\end{tabular}
\vspace{0.1in}
\caption{Indicator building model characteristics and associated probable building faults.}
\label{tbl:fault}
\end{table}

\subsubsection{Parameter relationship with building faults}
Heating slope $\gamma_{heat}$ and heating balance point temperature $T^{heat}$ are the two parameters that enable our model to interpret the heating inefficiencies of a home. 
Buildings with high $\gamma^{heat}$  lose heat at a higher rate, which in turn affects heating unit usage (i.e., consumes more power) to compensate for the high loss rate. A high energy loss rate can be attributed to poor building insulation, air leakages, or inefficient or heating unit. Separately, heating balance point temperature also indicates inefficiencies in the heating component of a home. A high balance point temperature suggests two possible inefficiencies: (i) high thermostat set-point temperature\footnote{Set point temperature and balance point temperature have a linear relationship} and (ii) poor building insulation. If the set-point temperature is high during winters, heating units turn on more frequently to maintain the indoor temperature at set-point. In contrast, if building insulation is poor, more heat is lost through the building envelope. Thus, heating units will be turned on frequently to sustain the high heating balance point temperature. Similarly, we can interpret the cooling slopes $\gamma^{cool}$  and cooling balance point temperature, which points to inefficiencies in cooling units or building envelope. 

Homes with high $E^{base}$ indicate high appliance usage or inefficient appliances.  In such homes, energy retrofits may not help reduce energy consumption.  However, these homes may benefit from replacing old appliances (water heater, dryer) with newer energy star rated ones. 
We summarize the association between probable causes of building faults and model parameter in Table~\ref{tbl:fault}. 

\subsubsection{Inefficient Home Analysis Algorithm}
 
\begin{algorithm}[t]
\caption{Identify Inefficient Homes Algorithm}\label{alg:outlier1}
\begin{algorithmic}[1]\\
{\bf Inputs}: Sensitivity ($\tau$), buildings ($B$)
\Procedure{FindInefficientHomesCohort}{$\tau$, $B$}
\State count = \{\}; homes = \{\}
\For {$p$  in $[\gamma^{heat}, \gamma^{cool}, E^{base}]$}
\For {($b1$, $b2$) $\gets$ $\Perm{|B|}{2}$} $\quad$// all-pairs permutation
\If {$F_p(b1) \succeq_{2} F_p(b2)$ }
\State count[$p$, $b1$] +=1
\EndIf
\EndFor
\For {$b$ $\gets$ $B$}
homes[$b$][$p$] = count[$p$, $b$] $\ge$  $\tau$
\EndFor
\EndFor
\For {$b$ $\gets$ $B$}
homes[$b$][$T^{heat}$] = $T_{b}^{heat}> 70^{\circ}F$
\EndFor
\For {$b$ $\gets$ $B$}
homes[$b$][$T^{cool}$] = $T_{b}^{cool}< 55^{\circ}F$
\EndFor
\State \Return homes
\EndProcedure
\end{algorithmic}
\begin{algorithmic}[1]\\
{\bf Inputs}: Candidate Building ($h$), Location ($L$), Attribute($A$), Cohort Size (S)
\Procedure{FindInefficientHomesRegion}{$h$, $L$, $A$, $S$}
\State $\hat{\gamma}^{heat}, \hat{\gamma}^{cool}, \hat{E}^{base}$ = getPeerCohortDistribution($L$, $A$, $S$)
\State home = \{\}
\State home[$\gamma^{heat}$] = $h[{\gamma}^{heat}] \succeq_{2}  \hat{\gamma}^{heat}$ 
\State home[$\gamma^{cool}$]  = $h[{\gamma}^{cool}] \succeq_{2}  \hat{\gamma}^{cool}$ 
\State home[$\hat{E}^{base}$]  = $h[{E}^{base}] \succeq_{2}   \hat{E}^{base}$
\State home[$T^{heat}$] = $h[T^{heat}]> 70^{\circ}F$
\State home[$T^{cool}$] = $h[T^{cool}]< 55^{\circ}F$
\State \Return home
\EndProcedure
\end{algorithmic}
\end{algorithm}
 
 We present the pseudo-code to determine inefficient buildings in Algorithm~\ref{alg:outlier1}.
Depending on the execution mode, we can use our algorithm to find inefficient homes within a cohort or identify whether a candidate home is inefficient. Below, we outline both scenarios.
\\ 
 \noindent  {\bf Identify Inefficient Homes within a Cohort: }
In this scenario, we identify homes that are inefficient homes within a cohort. To do so, we first use the partially ordered set of buildings to determine the outliers for each parameter. 
To determine outliers, note that the energy usage of an inefficient home would be high. Thus, the building parameter distribution of an inefficient home will tend to be \emph{stochastically dominant} with respect to others in their peer group. 
However, among inefficient homes, the building parameter distribution may be similar, and thus their distributions may not be stochastically dominant to one another.
Similarly, within energy efficient homes this distinction of dominance may not be apparent, as their distribution may be identical to one another. We use this insight to define a building as \emph{inefficient} in a given model parameter, if it is stochastically dominant compared to a majority of the homes within its cohort.  For instance, if a building's heating parameter distribution $F_{\hat{\gamma}^{heat}}$ is dominant across more than $\tau$\% of the buildings, we conclude that the building is inefficient and has a \emph{high} heating slope. 
Here, $\tau$ is the sensitivity threshold for \texttt{WattScale} and provides the flexibility to control the number of inefficient homes. 
The higher the threshold value, the higher the possibility of identifying an inefficient home.  
For all experiments, we chose this to be 75\%.
Thus, for each parameter, we determine whether a building is inefficient if its distribution is dominant beyond a certain threshold. 
We use a balance point threshold to determine buildings with high balance point temperature. We flag buildings as inefficient if the mean value obtained after inference for heating (or cooling) balance point temperature $T^{heat}$ (or $T^{cool}$) is greater than (less than) specific heating (or cooling) balance point threshold 70\textdegree F (55\textdegree F) ---  a common choice employed by expert auditors. However, these values can also be provided as parameters to the algorithm. The pseudo-code to determine inefficient homes within  a cohort is presented in Algorithm~\ref{alg:outlier1}. 
\\
\noindent {\bf Identify Inefficient Home within a Region:} To identify whether a candidate building is inefficient, we use their location information to first create a cohort for comparison. The difference between the previous scenario is that, here, the cohort is not provided in advance. We create the cohort based on the region and the attributes of the candidate building. Further, unlike the previous scenario, where the task is to identify all homes that are inefficient within a cohort, it requires performing an all-pairs comparison within the cohort. In this case, we only have to compare the candidate building against the region's building parameter distribution to examine whether it is inefficient. This approach is illustrated in Algorithm~\ref{alg:outlier1}. Our approach finds all candidate buildings in a location $L$ that meets the criteria specified in attribute $A$. Depending on the size of the cohort, our approach expands on the search over a region until sufficient buildings are identified to create the cohort. Once the peer cohort is created, we create and the building distribution parameters of the cohort namely the heating slope $\hat{\gamma}^{heat}$, cooling slope $\hat{\gamma}^{cool}$, and the base load $\hat{E}^{base}$ of the region. These parameters are then used to compare against the candidate building to identify any inefficiencies. For instance, if the candidate building's cooling slope is stochastically dominant compared to the region-wide cooling slope parameter, then we indicate it to be inefficient. 

%


\subsubsection{Root Cause Analysis}
As noted earlier, each parameter in the building model affects an energy component defined in ~(\ref{eqn:model}). 
Any irregularity in the building parameter, in comparison to its peer group or the region, points to possible inefficiency in the said energy component. We outline our pseudo-code for finding root cause in Algorithm~\ref{alg:outlier3}. First, we create a mapping of indicators of deviations in building model parameters to possible faults using Table~\ref{tbl:fault}. We provide the mapping as an input to our algorithm. Next, we associate a fault to a home if it was flagged inefficient for the given parameter $p$. For instance, if a home is flagged as high base load, we say that the home has inefficient appliances. Similarly, an inefficient home with high heating slope is assigned faults related to heating inefficiencies. We then generate a report of the list of potential faults in a given home.

\begin{algorithm}[t]
\caption{Fault Analysis Algorithm}\label{alg:outlier3}
\begin{algorithmic}[1]\\
{\bf Inputs}: building ($h$), parameters ($P$), fault\_map ($M$)
\Procedure {GetRootCause}{$h$, $P$, $M$}
\State faults = []
\For {$p$ $\gets$ $P$}
\If {h[$p$]}
\State faults += $M[p]$ $\quad$ // append list
\EndIf
\EndFor
\State \Return faults
\EndProcedure
\end{algorithmic}
\end{algorithm}

\section{Implementation}
\label{sec:implementation}


\texttt{WattScale} is split into two components --- (i) a Unix-like command line tool~\footnote{We have publicly released the code and the tool. \url{http://bit.ly/2nU7kA5}} 
that uses PyStan, a statistical modeling library, to implement our bayesian model, and 
(ii) a web-based application interface implemented using Django framework for interacting with the command line tool. 
Users can interact with either component to determine likely reasons of inefficiency of an individual building or a group of buildings. 
 

\begin{figure}[t]
\centering
\begin{tabular}{cc}
\includegraphics[width=2.4in]{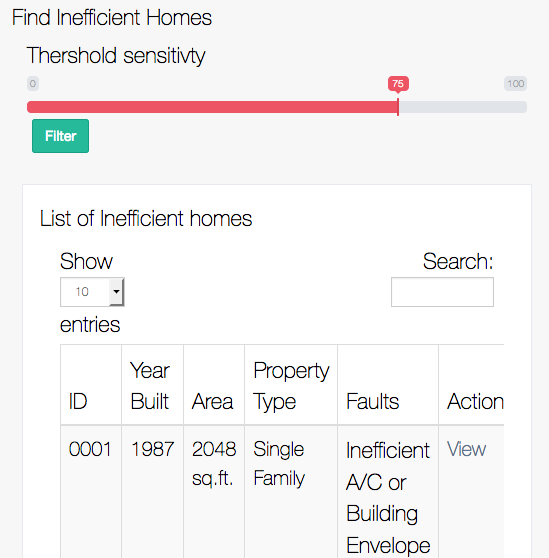} &
\includegraphics[width=2.4in]{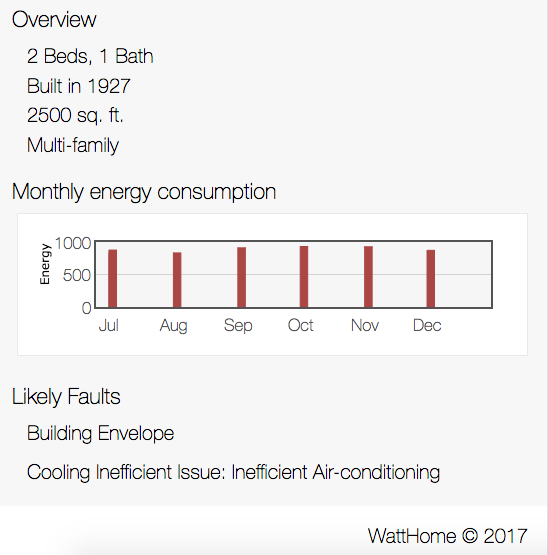}  \\
(a) Find Inefficient homes & (b) Inefficiency Report \\
\end{tabular}
\caption{Screenshot of our implementation of \texttt{WattScale}.}
\label{fig:app}
\end{figure}

To determine the inefficiencies in a single building (i.e., region-based execution mode), we provide users an interface to upload their Green Button friendly format energy usage information~\cite{greenbutton}. 
The Green Button initiative provides energy consumers access to their energy consumption data collected from their smart meters. Since many utility companies widely support the Green Button format, this enables our service to be used by millions of consumers in the US. When users upload their Green Button data, along with building information (such as zip code, year built, etc.), \texttt{WattScale} creates a custom bayesian model of the home using the local weather data and the details provided by the user. 
The weather data of a nearby airport is used as a proxy for local weather conditions, and \texttt{WattScale} periodically fetches and updates this data from public APIs.  
Further, we use the location data to create a cohort group that matches the attribute of the building provided. For instance, if the user's building is a single family home that was built in the year 1940, our algorithm uses this information to create a peer cohort having similar features, that is, single family homes built around 1940 under similar climate conditions, to enable a fair comparison. We expand our search space, using R-tree based data structure, to identify additional homes if there are not sufficient homes in a given region that match the filter criteria. Next, we create a building parameter distribution of the cohort and compare it with the candidate home provided by the user to determine inefficiencies. We then highlight the likely faults in the home.


\texttt{WattScale} can also identify inefficient homes within a group of buildings (i.e., individual execution mode). This mode is useful for utility companies that have access to energy data from several homes in a region to identify a set of homes that are energy inefficient. In this mode, a user uploads the energy information and building attributes for a group of buildings. 
Here, we assume that the weather conditions are similar for all the input buildings. 
As again, we use the location of the building to retrieve the local weather data and build a custom bayesian model of the home. We also use the building attributes to create custom peer groups within the set of buildings. Next, users provide a sensitivity threshold that is used to create a  partially ordered set of inefficient homes. 
As utility companies may have a limited audit budget to manually inspect homes, the threshold provides a user the flexibility to control the list of least efficient home. 
Figure~\ref{fig:app}(a) shows how users can adjust the sensitivity parameter to get inefficient homes. 
Finally, our \texttt{WattScale} generates a report listing inefficient homes and their likely faults. Figure~\ref{fig:app}(b) shows the inefficiency report for a single home listing likely faults.

\section{Experimental Validation}
\label{sec:evaluation}
We first validate our model estimates against ground truth data from three cities and evaluate its efficacy. For each of these datasets, we convert the heating fuel type usage to kWh equivalent. 


\subsection{Dataset Description} 
\subsubsection{Dataset 1: Dataport (Austin, Texas)} 

Our first dataset contains energy consumption information from homes located in Austin, Texas from the Dataport Research Program~\cite{dataport}.
The dataset contains energy breakdown at an appliance level, which serves as ground truth to understand how our approach disaggregates energy components. 
We select a subset of homes (163 in total) from this dataset having HVAC, baseload appliances along with the total energy usage information. 
Since most homes enrolled in the Dataport research program are energy-conscious homeowners, and have energy efficient homes, we use this dataset only for validating our energy disaggregation process. 

\subsubsection{Dataset 2: Utility smart meter data (New England)}

\begin{table}[]
\centering
\begin{tabular}{|c|c|c|c|}
\hline
\textbf{Charactersitics} & \textbf{Dataset 1} & \textbf{Dataset 2} & \textbf{Dataset 3} \\ \hline
\# of Homes              & 163                & 10,107      & 32           \\ \hline
Duration                 & 2013          & 2015     & 2014-15            \\ \hline
Built Area Range (sq.ft.)         & 758-6516        & 250-10,000  & 1030-4673         \\ \hline
Year Built Range         & 1912-2014          & 1760-2013 & 1910-2004          \\ \hline
Location         & Austin, TX          & A city in New England & Boulder, CO         \\ \hline
\end{tabular}
\vspace{0.1in}
\caption{Key characteristics of Dataport and New England-based utility smart meter dataset}
\label{tbl:dataset}
\end{table}

This dataset contains smart meter data for 10,107 homes  from a  small city in the New England region of the United States~\cite{iyengar2016analyzing}. 
The dataset has energy usage (in kWh) from both electricity and gas meters. 
Each home may have more than one smart meter --- such as a meter to report gas usage and another to report electricity usage. 
For homes with multiple meters (gas and electric), we combine their energy usage to determine the building's daily energy consumption for an entire year (2015). 
Apart from energy usage, 
the dataset also contains real estate
information that includes building's size, the number of rooms, bedrooms, property type (single family, apartment, etc.). We also have manual audit reports for some of the homes. 
We use this as our ground truth data for validating our approach. 
Further, we have weather information of the city containing average daily outdoor temperature. 

\subsubsection{Dataset 3: Dataport (Boulder, Colorado)} 

Our third dataset contains energy consumption information from homes located in Boulder, Colorado from the Dataport Research Program~\cite{dataport}.
Similar to dataset 1, this one contains energy breakdown at an appliance level.
We select a subset of homes (32 in total) from this dataset having several appliances along with the total energy usage information for a period of one year. We will use this dataset to validate the performance of WattScale in identifying inefficient homes using the distribution of building model parameters for a region. 
We summarize the characteristics of all three datasets in Table~\ref{tbl:dataset}.

\subsection{Energy Split Validation}


\begin{figure*}[t]
   \centering
\minipage{0.43\textwidth}
  \includegraphics[width=\linewidth]{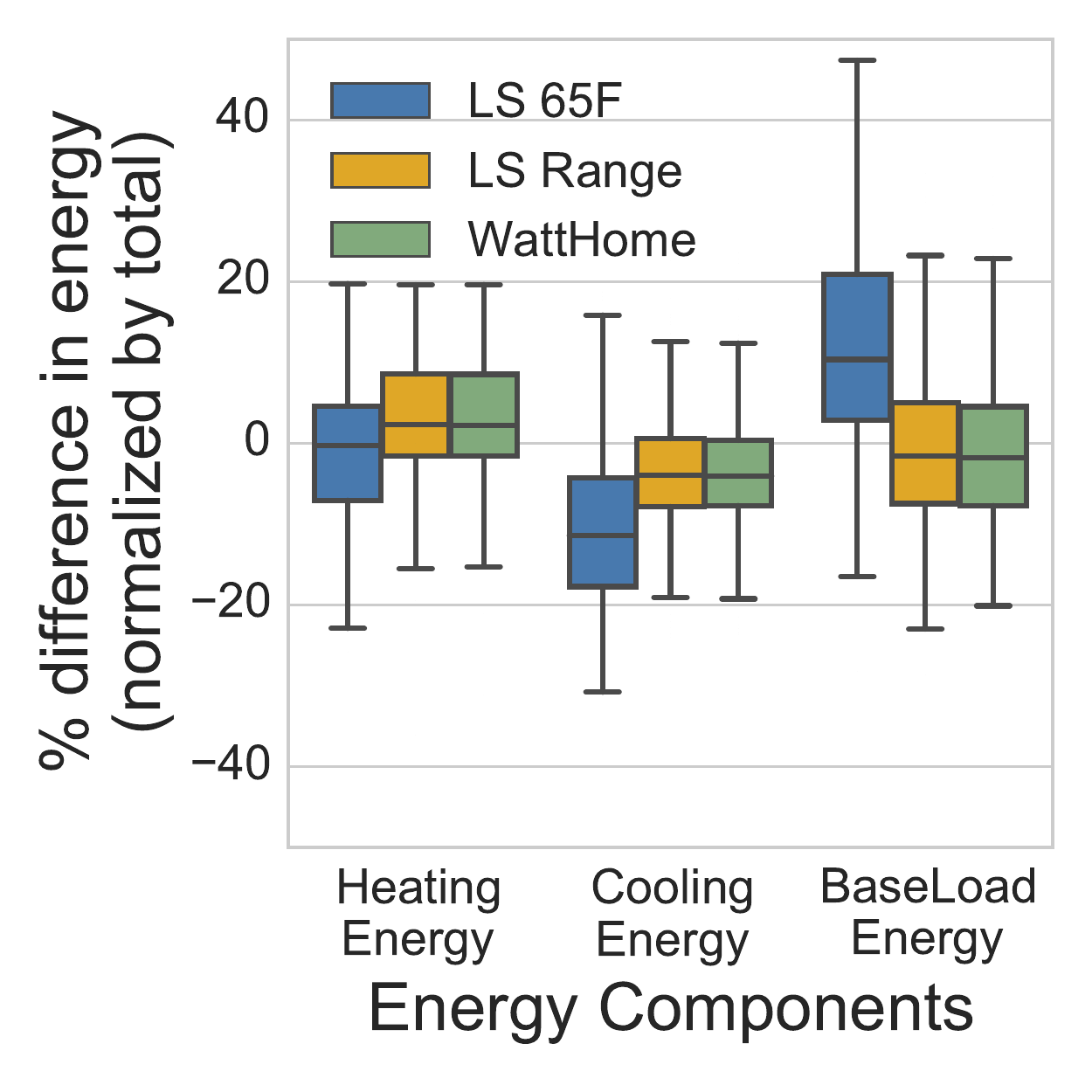}
  \caption{Validation of energy split using the two baselines and our model.}
 \label{fig:pecan_split}
\endminipage \hfill
\minipage{0.43\textwidth}
  \includegraphics[width=\linewidth]{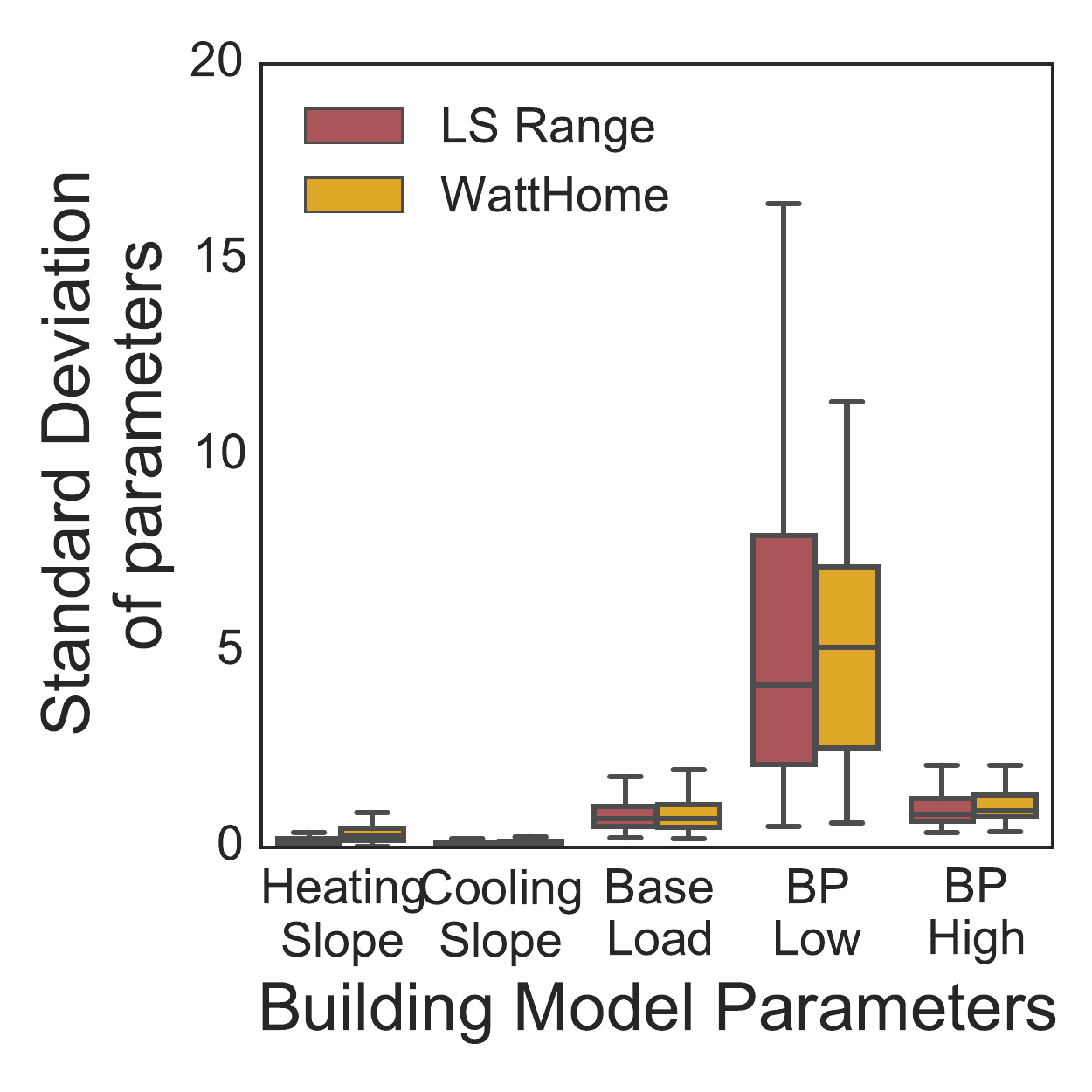}
  \caption{Comparison of the standard deviation of parameters.}
   \label{fig:pecan_sd}
\endminipage\hfill
\end{figure*}

%

We now validate the efficacy of our model in disaggregating the overall energy usage into distinct energy components, i.e., heating, cooling, and baseload. For this experiment, we restrict our analysis to the 163 homes from the Dataport (Austin) dataset.

We compare our technique with two baseline techniques (\emph{LS~65F} and \emph{LS Range}), commonly used in prior work, which use the degree-days model to provide point estimates of the individual building model parameters. 
Our first baseline technique, \emph{LS 65F}, estimates the three building energy parameters ($\gamma^{heat}$, $\gamma^{cool}$, $\sigma$, $E^{base}$) using least-squares fit and assumes the balance point temperature to be constant (65$^\circ$F). This is a widely used approach by energy practitioners around the US and recommended by official bodies such as ASHRAE~\cite{ashrae2013fundamentals}. Our second baseline technique, \emph{LS Range}, estimates all the five building energy parameters ($\gamma^{heat}$, $\gamma^{cool}$, $T^{heat}$, $T^{cool}$, and $E^{base}$) using the least-squares fit. 
Unlike the baseline approaches, \texttt{WattScale} estimates the parameter distribution and thus to compare we use the mean of the posterior distribution of the parameters to get the fixed proportion of the energy splits. 

Figure~\ref{fig:pecan_split} shows the distribution of percentage difference in the energy usage with the ground truth for each energy component. 
While \emph{LS Range} and \emph{\texttt{WattScale}} have median error of $\approx$-1.6\%, \emph{LS 65F} have a median error of 10\% for baseload energy. 
Unlike \emph{LS 65F}, \emph{LS Range} and \emph{\texttt{WattScale}} do not assume a constant balance point temperature and thus have lower error. 
Figure~\ref{fig:pecan_sd} compares the standard deviation of the building parameters from the two approaches. 
In \emph{\texttt{WattScale}}, the standard deviations are obtained from the parameter posterior distributions. Whereas, in case of \emph{LS Range}, the standard deviations are calculated from the covariance matrix outputted by the least-squares routine. 
While the results for the four parameters are similar, the spread of standard deviation for the lower balance point is much smaller in \emph{\texttt{WattScale}} compared to \emph{LS Range}. Thus, \emph{\texttt{WattScale}} provides an equivalent or tighter bound compared to \emph{LS Range}. 
\\
\textbf{Summary}: \emph{Fixed parameters provide poor estimate of the building parameter. \texttt{WattScale} provides  lower error and tighter parameter estimates  compared to other baseline techniques.}

\subsection{Faulty Homes Validation}

We now examine the accuracy of our model in reporting homes with likely faults. We ran our algorithm on all homes in the New England dataset to generate a list of outlier homes for each of the parameter and then compare our results 
with findings from manual energy audits (ground truth).
Since manual audit reports contain faults related to building envelope and HVAC devices only, we only report these results and inefficiencies arising from base energy usage and faulty set points were not analyzed. 

To determine the accuracy, we compare an inefficient building's parameter to the audit report conducted in the past and verify whether it has any building faults.
The audit reports were manually compiled by an expert on-field auditor identifying and suggesting energy efficiency improvement measures. 
We find that \texttt{WattScale} reported 59 homes with building envelope faults, out of which 56 buildings were in the audit report, an accuracy of 95\%. Moreover,
we find that 46 of the 56 homes with building envelope faults also had faulty HVAC systems. 
\\
 \textbf{Summary}: \emph{\texttt{WattScale} identified parameter related faults in a building with high accuracy. In particular, our approach correctly identified 95\% of the homes that were flagged by  expert auditors as having either faulty building envelope or HVAC systems.}

\section{Case study: Identifying Inefficient Homes In A City }
\label{sec:casestudy}

We conduct a case study on the New England dataset to determine the least efficient residential buildings in the city using the individual execution mode.
In particular, we seek to gain insights on the following questions:
(i) What percentage of the homes are energy inefficient? 
(ii) Which groups of homes are the most energy inefficient? 
(iii) What are the most common causes of energy inefficiency? 
We first provide a brief analysis of the distribution of the energy split. 

\subsection{Energy Split Distribution Analysis}

\begin{figure*}[t]
\centering
\begin{tabular}{ccc}
\includegraphics[width=3in]{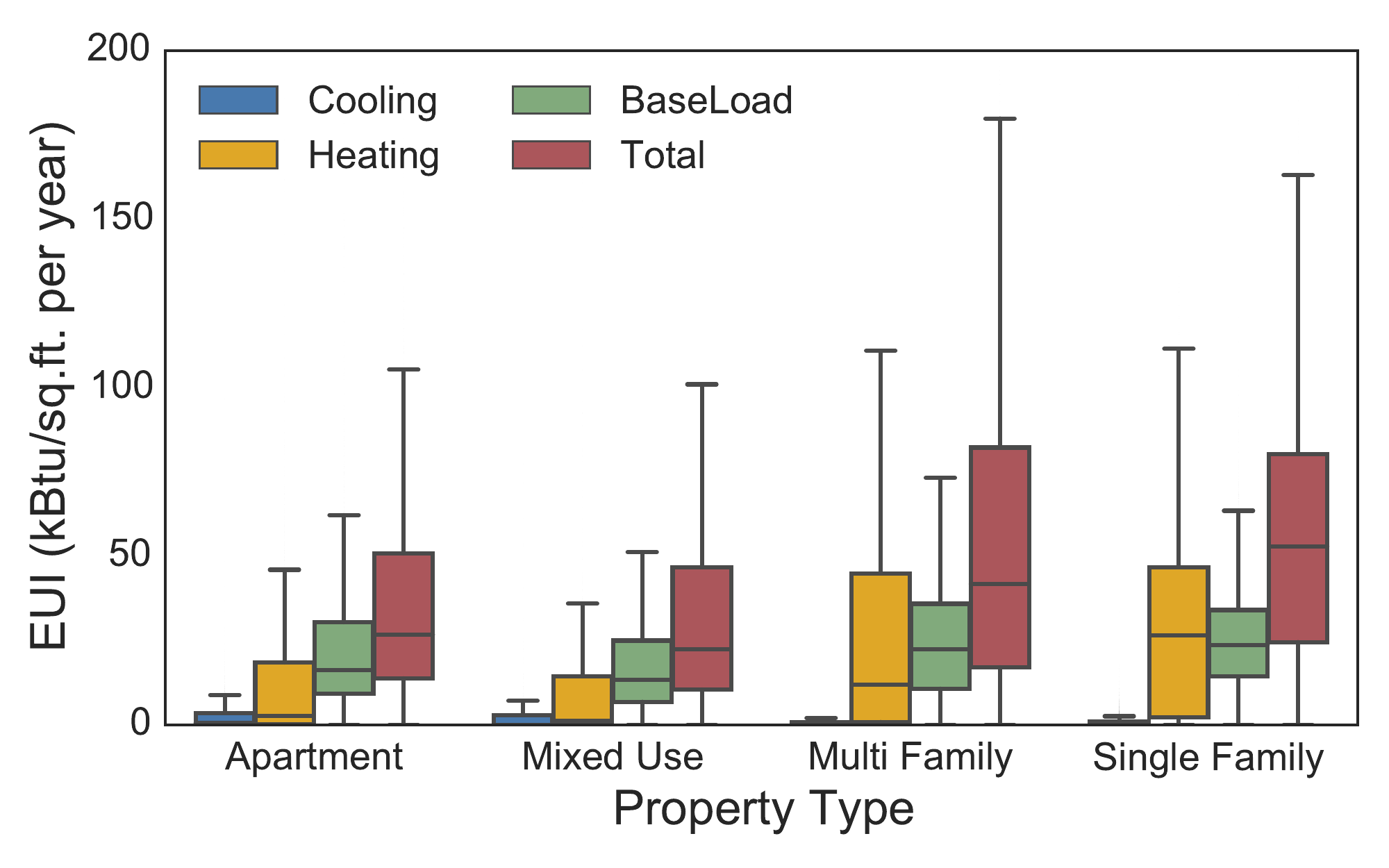} &\\ (a) Total Energy Split \\
\includegraphics[width=3in]{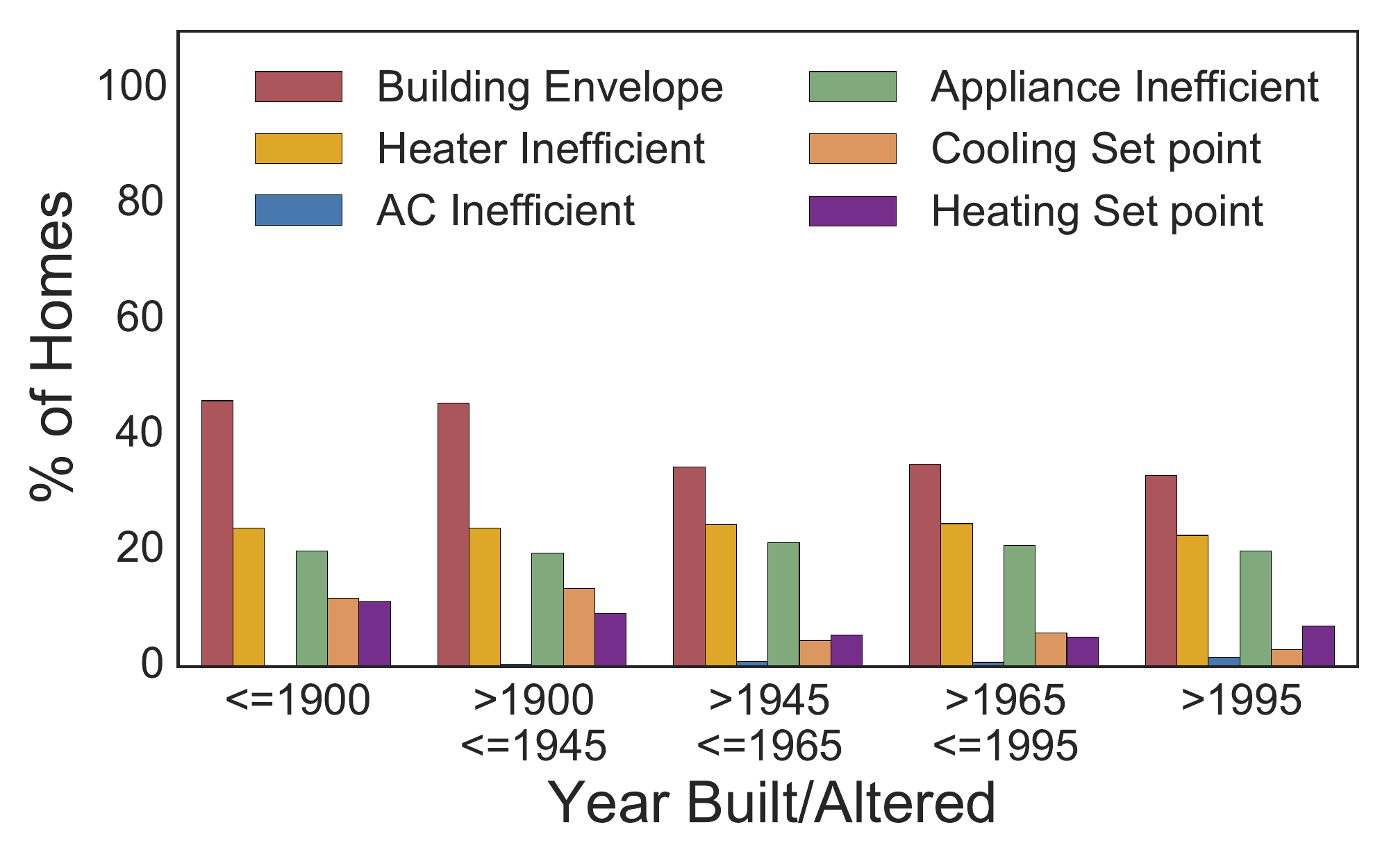} &\\ (b) By Building Age \\
\includegraphics[width=3in]{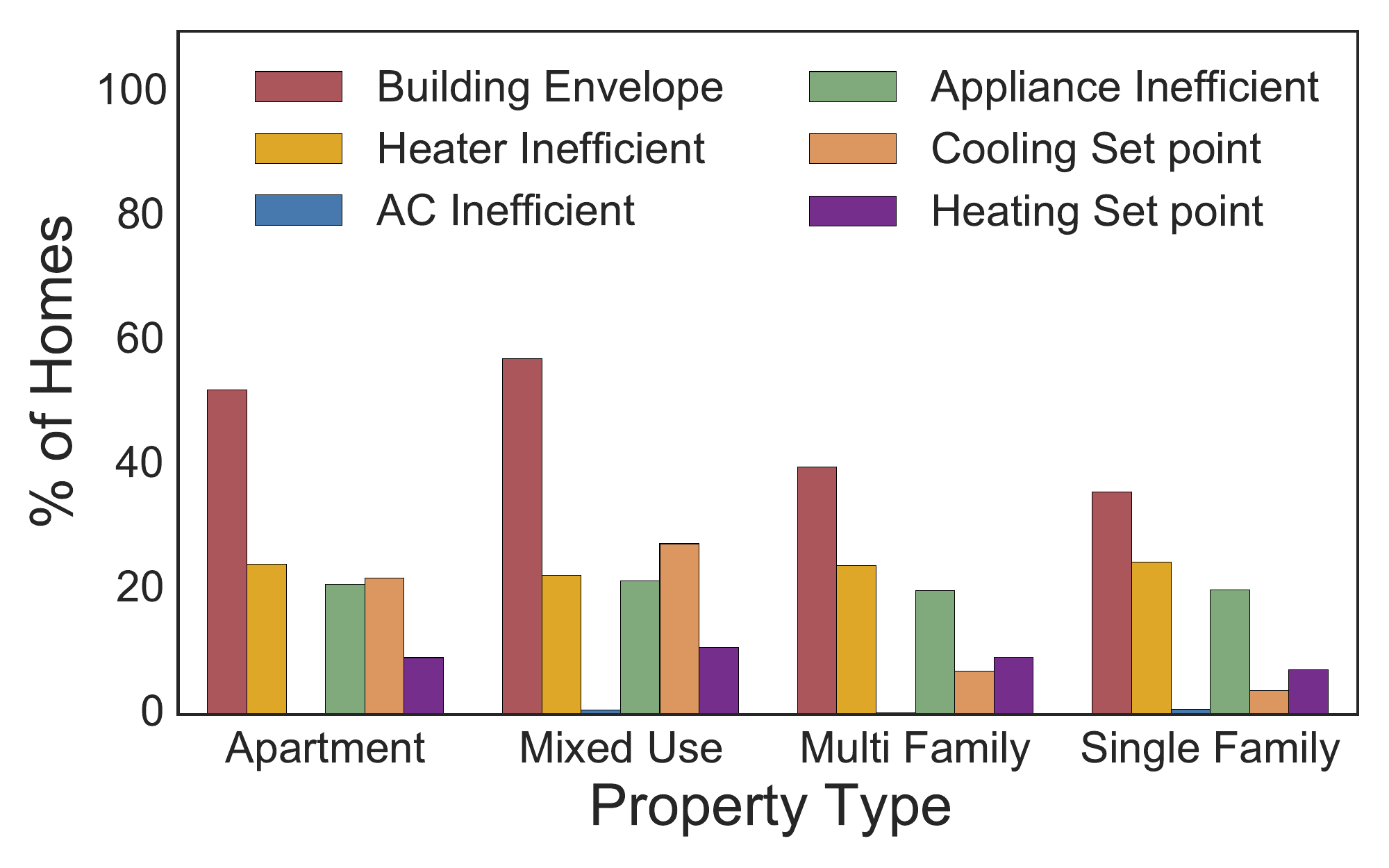}  \\
 (c) By Property Type \\
\end{tabular}
\caption{(a) Disaggregated energy usage for all homes. (b) and (c) Possible fault types in different building groups.}
\label{fig:faultage}
\end{figure*}

To get the fixed proportion of the energy split, we use the mean of the posterior estimates to compute the disaggregated energy usage i.e. heating, cooling and base load components. 
To compare the energy components, we compute the \emph{Energy Usage Intensity} (EUI), by normalizing the energy component with the building's built area.
Figure~\ref{fig:faultage}(a) shows the heating, cooling, base load and total EUI distribution grouped by property type across all homes.  
The figure shows that the base load is the highest component of energy usage in most Mixed Use and Apartment property types followed by heating and cooling. However, for Single family homes, the heating cost is usually higher.
The high base load can be attributed to lighting, water heating, and other appliances. 
Further, since the New England region has more winter days, homes require more heating, and thus expected to have a higher heating energy footprint compared to cooling. 
In particular, the average heating energy required is almost 20$\times$ that of average cooling energy. 
We also observe that the normalized total energy usage of single and multi family homes is the highest --- presumably due to more number of appliances.
The median energy EUI of the Single family home is $\approx$53 kBtu/sq.ft. (1 kW=3.412kBtu), which is almost twice that of Apartment homes ($\approx$26.8 kBtu/sq.ft.). 

\textbf{Observation}: \emph{Heating energy consumption is 20$\times$ that of cooling energy on an average. Energy consumption among Single and Multi family homes is much higher than Apartment or Mixed use homes.}

\subsection{Efficiency Analysis}
In this section, we analyze the results of our approach on the utility company's  dataset described earlier. 
We created peer groups  to identify inefficient homes in their respective cohort. 
To do so, we used three building attributes (property type, age, and area), which created 120 peer groups in total. 
Among these peer groups, we discarded groups with less than 20 homes, as it didn't have enough population size for a meaningful analysis. 
In all, 67 peer groups containing a total of 186 homes were discarded. Below, we present our analysis on the remaining 9,921 homes.

\subsubsection{Identifying inefficient homes}

\begin{table}[t]
\begin{tabular}{|c|c|c|c|}\hline
Heating & Cooling& Base load&Overall\\
Outliers & Outliers& Outliers& Outliers\\\hline
3162 &1033 &2016 & 5079 \\\hline
\end{tabular}
\vspace{0.1in}
\caption{Summary of all inefficient homes in the data set.}\label{tbl:newout}
\end{table}%

We examine the number of homes that are flagged as inefficient for each of the energy components using our approach. 
Table~\ref{tbl:newout} shows the summary of inefficient homes
across all peer groups. We note that a home may have multiple inefficiencies, such as inefficient heating and high base load 
and thus may be inefficient in several of the energy components. 
Our results show that the overall percentage of inefficient homes across all residential homes is 50.25\%.
Further, almost 62.25\% of all inefficient homes have either inefficient heater or poor building envelope, and 4144 homes have either inefficient heating or cooling~\footnote{The number of outliers found will vary due to changes in geographic regions, prevalent building codes, age of the building, property type, etc. We are not making any claims regarding the generality of the final results of the analysis across geographies. However, the analysis itself is quite general and can be applied to data from any part of the world as the main categories of building faults do not change from region to region.}.
\\
\textbf{Observation}: \emph{More than half of the buildings in our dataset are likely to be energy inefficient, of which almost 62.25\% homes have  inefficient heating as a probable cause.}

\subsubsection{Identifying faults in inefficient homes}

We now analyze the cause for inefficiency in these inefficient homes. 
Figure~\ref{fig:faultage}(b) shows the percentage of inefficient  homes within each building age group across all faults. 
Note that a home may have multiple faults. 
 We observe that the building envelope fault is the major cause of inefficiency, 
followed by inefficiency in heaters and other base load appliances. Across all age groups, nearly 41\% of the homes have building envelope faults, 
while 23.73\% and 0.51\% homes have heating and cooling system faults respectively. 
The figure also shows that some homes might have set point faults. 
In particular, 18.06\% of the homes have issues with either high heating or low cooling set point temperature. 
These faults indicate likely issues with thermostat setting. 
Adjusting the thermostat set point temperature in these home may likely improve its efficiency. As shown, homes built/altered before 1945 have a higher proportion of inefficient homes. However, the percentage difference with other age groups is <15\%. 


Figure~\ref{fig:faultage}(c) shows the percentage of inefficient homes within each building property type and faults. 
We observe that the building envelope faults are the most common faults across all building types. 
Further, we find that except for HVAC appliance related faults, 
mixed use property type has the highest percentage of inefficiency in the remaining fault categories.
After mixed use property type, apartments tend to have a higher percentage of inefficient homes
followed by multi family and single family property types. 
\\
\textbf{Observation}: \emph{Building envelope faults is one of the major cause for inefficiency and present in nearly 41\% of homes. 
However, 18.06\% of homes have thermostat set point faults. Changing their set-point may likely improve efficiency in these homes. }

\subsubsection{Neighborhood Analysis}
 \begin{figure}[t]
\includegraphics[width=3in]{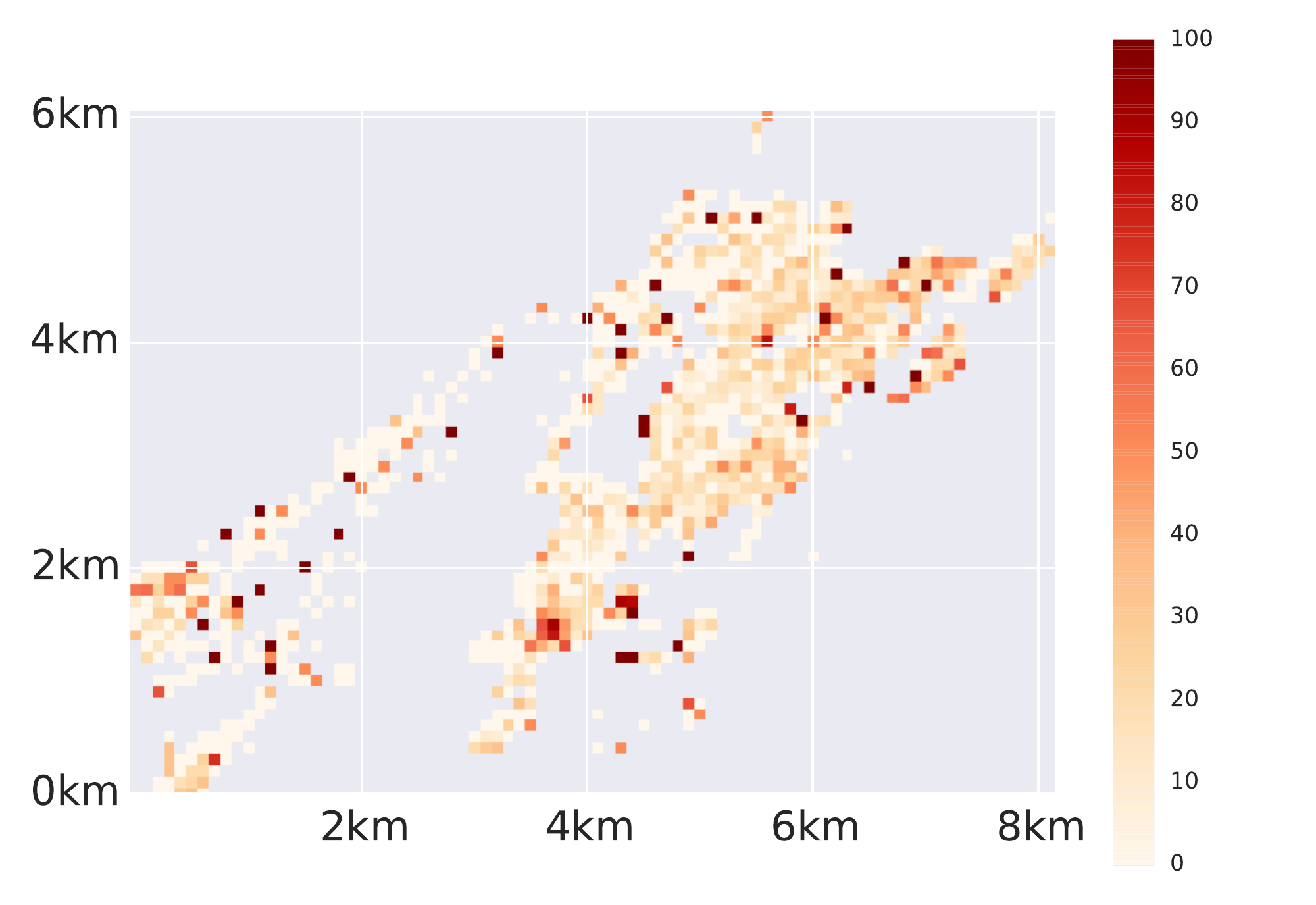}
\caption{Spatial distribution of inefficient homes in the city.}
\label{fig:geospatial}
\end{figure}

We plot inefficient homes spatially to observe whether inefficient homes are clustered together. 
To anonymize the data, we partition the map into 5166 grids of size 100$\times$100 meters. Further, we bucketize all homes in these grids
and report the percentage of inefficient homes  within each of them. 
Figure~\ref{fig:geospatial} shows the heat map of the percentage of homes that are inefficient in each grid. The gray sections in the figure are uninhabited areas with no buildings. 
The light colored patches are areas with few or no inefficient buildings, while the darker colored areas reveal a higher proportion of inefficient buildings. 
As seen in the figure, most inefficient homes are co-located.  
In particular, we find that just 100 grids (=1 sq. km. area) out of the overall 51.66~sq.~km.~area has more than 50\% of all inefficient homes. 
\\
\textbf{Observation}: \emph{Most inefficient homes are co-located. In particular, 50\% of all inefficient homes lie in 1 sq. km. area.}

We summarize the result in Table~\ref{tbl:newout}.
In percentage terms, among the mixed use peer groups 33.33\% of the homes are inefficient.  
While, in the case of single family peer groups, the fraction of inefficient homes is only 12.74\%. 
However, in absolute values, single family property type has the highest number of inefficient homes (575 homes) followed by apartments (558 homes). 
Since most of the apartment homes belong to the older age group i.e. buildings built before 1945, these groups can be likely candidate targets. 
We also observe that in some age groups, there were few outliers, which can be attributed to fewer homes in these groups. 
\\
\textbf{Observation}: \emph{Newer homes are more energy efficient than older ones. Homes built before 1945 represent $\approx$72\% of the total outliers.}

\section{Case study: Identifying Inefficient Homes Anywhere in the US}
\label{sec:casestudy_extra}

We present another case study on the Dataport (Boulder) dataset to validate the energy efficiency results from our scalable region-based execution mode with the results obtained from the individual execution mode. To get the distribution of building parameters of a region, we use the publicly available Building Performance Database (BPD)~\cite{bpd}. BPD is the United States' largest dataset containing energy-related information of commercial and residential buildings. 

  \begin{figure}[t]
\includegraphics[width=3.7in]{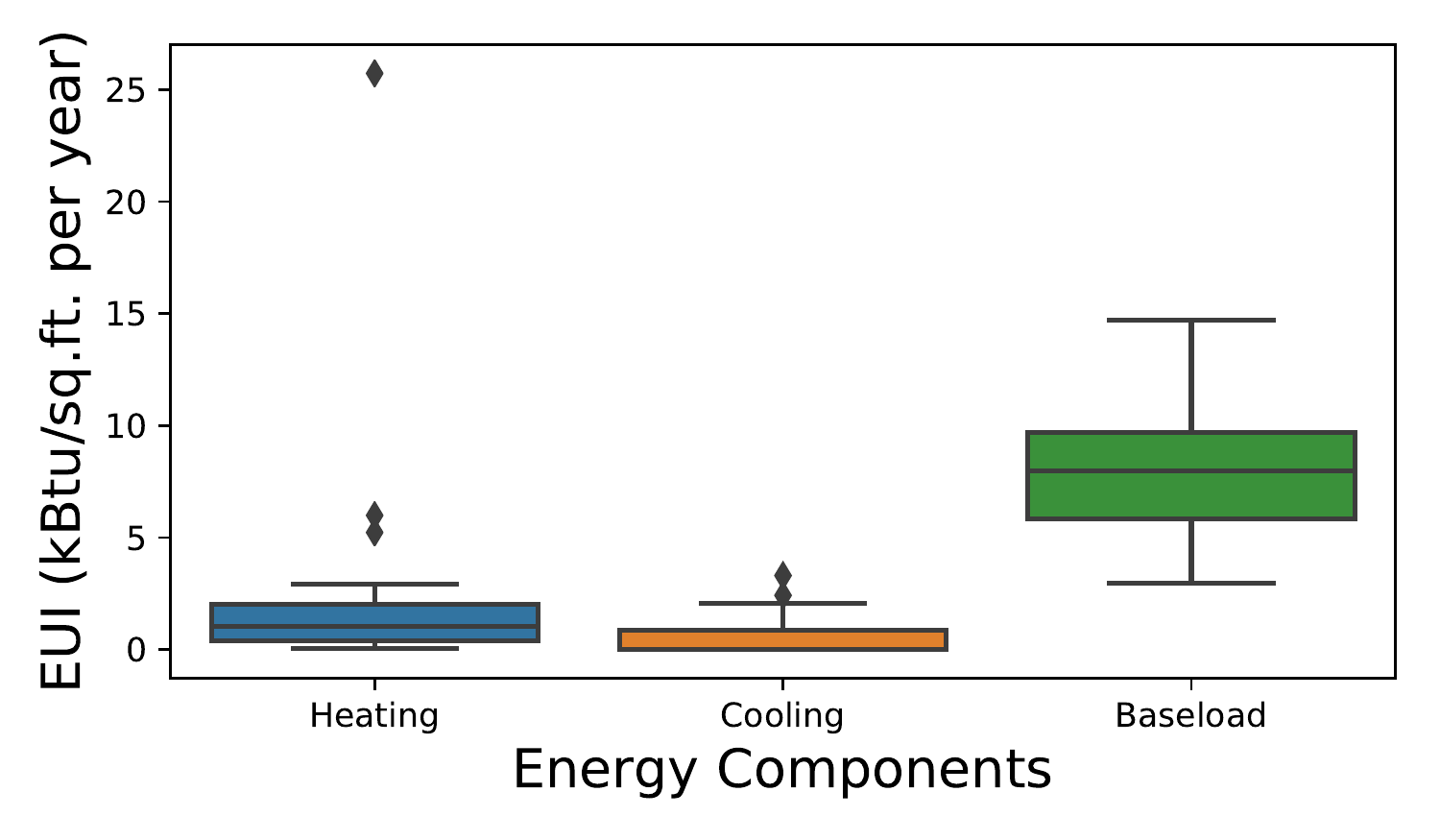}
\vspace{-0.1in}
\caption{Energy Components of the single family homes in Boulder (Dataport dataset)}
\label{fig:base_load_heat}
\end{figure}

\subsection{Energy Split Distribution Analysis}

Once again to get the fixed proportion of the energy components, we use the mean of the posterior estimates of the building model parameters. Figure~\ref{fig:base_load_heat} shows the heating, cooling, and the base load EUI's distribution across all the 32 homes. In this dataset, baseload energy component dominates energy consumed for heating and cooling. The average daily energy consumed to run non-HVAC appliances is almost 3.88x and 15.53x that of average heating and cooling energy respectively. In fact, 15 of the 32 homes have zero cooling needs.

\textbf{Observation}: \emph{Baseload is the dominant component of energy usage in Dataport (Boulder) dataset. Cooling is a very small proportion of energy used among the single family households in Boulder, CO.}

\subsection{Energy Efficiency Analysis}
 
  \begin{figure}[t]
\includegraphics[width=5in]{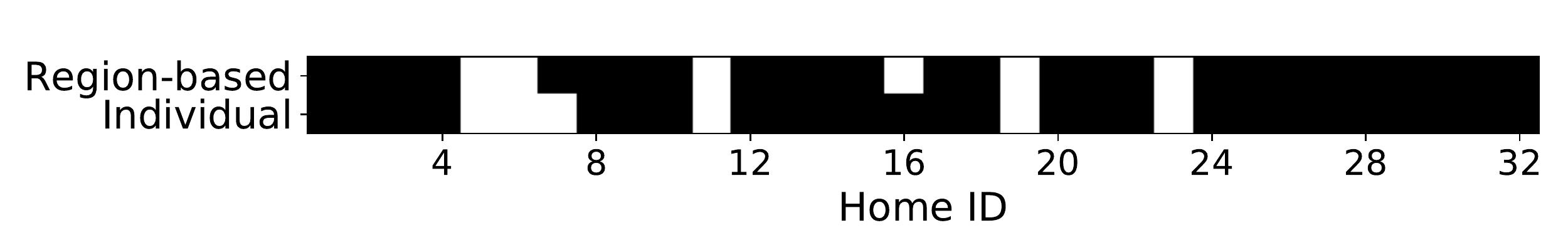}
\vspace{-0.1in}
\caption{Heatmap showing results of outlier homes for baseload}
\label{fig:base_load_heat}
\end{figure}

  \begin{figure}[t]
\includegraphics[width=5in]{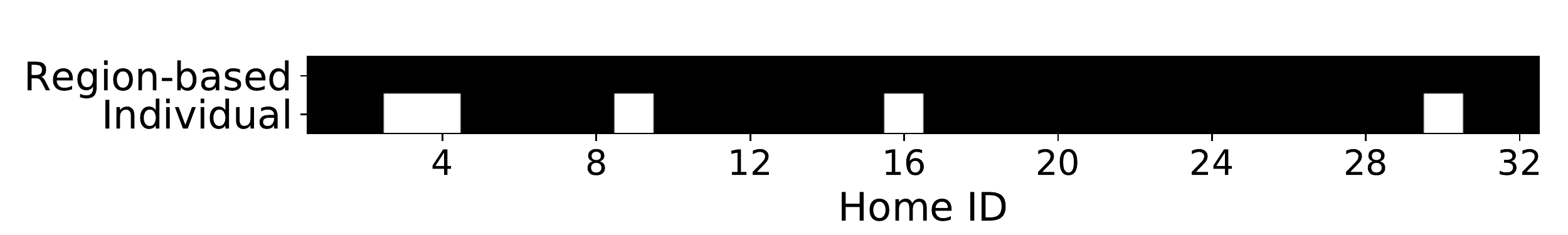}
\vspace{-0.1in}
\caption{Heatmap showing results of outlier homes for heating slope}
\label{fig:heat_slope_heat}
\end{figure}

We now compare the efficacy of the region-based mode in \texttt{WattScale} with the individual mode used to flag inefficient homes. The individual mode needs energy consumption data from several homes present in the same \textit{peer-group}. Whereas, in the region-based mode, we compute the building model parameters of all the homes in the region to identify the causes of inefficiency. 

Figure~\ref{fig:base_load_heat} shows the baseload outliers identified by the two modes. As shown, both modes discover 6 homes with excessive baseload energy usage. 5 of the 6 homes flagged by the two modes are common, pointing to significant agreement between them. Home IDs 7 and 16 were only identified by one of the two modes. 

Figure~\ref{fig:heat_slope_heat} shows the heating slope outliers identified by the two modes. Here, the region-based approach did not flag a single home. Whereas, the individual mode detected 5 out of the 32 homes to have a higher heating slope. This discrepancy exists as in the region-based mode as we compare the heating slope distribution of each home with the distribution learned from the various homes in BPD, a highly representative set of residential buildings in any region of the US. On the contrary, homes in the Dataport (Boulder) dataset consist of energy conscious households that have undergone energy audits.

\textbf{Observation}: \emph{ \texttt{WattScale} provides two execution modes to flag inefficient homes.  The region-based mode provides comparable performance to the individual mode, provided the dataset used to compute the model parameter distributions for the region is from a representative set of homes.}

\section{Discussion}
\label{sec:discussion}

With increasing penetration of smart meter data, building energy usage is easily accessible for a wide population of consumers. At the same time, weather and real-estate data has never been more readily accessible for major parts of the world. Since WattScale uses coarse-grained daily and annual energy consumption to create distribution for a building and region, respectively, we see enormous potential in applying our data-driven approach for various energy-efficiency related analytics. We note that distribution of building parameters for a region can be computed easily as several utilities provide typical load profiles of different sizes of residential homes they serve~\cite{bge,northwest,sdge}.
In this section, we briefly describe how our approach can benefit various stakeholders to improve the overall energy efficiency of buildings.



\textbf{Utility Companies:} As discussed earlier, our approach helps identify inefficient buildings within a cohort. This information when combined with geospatial data can reveal inefficient neighborhoods that can benefit from utility-scale energy awareness drives. Such energy awareness campaigns can foster better customer engagement and also improve the overall energy-efficiency of the locality. Further, based on the likely faults identified, special evidence-based policies can be designed to target inefficient groups and maximize its impact. 

\textbf{Policymakers and Government entities:} 
In the US, rebates and incentives are provided both at the federal~\cite{energysaver} and the state levels~\cite{vermont,california}. Policymakers can assess the impact of various subsidies and how it will impact the overall energy consumption. When combined with other information, such as census data, one can target subsidies to economically poor households. These households can benefit from government subsidies to not only improve their overall energy efficiency but also help save money. 

\textbf{Researchers:} Since our approach can be used beyond the city-scale for different regions, researchers can use our system to study the impact of pre- and post-retrofit modifications in a home and perform randomized tests (A/B Testing). Our tool can be used to create control groups based on various factors such as year built, area, fuel type that affect the efficiency of a building.  For example, if a county has incorporated a new energy policy for providing rebates/subsidies, we can assess the impact of the policy by comparing it to other counties. Our tool can also be used for longitudinal studies that record several measurements over multiple years. We can create a building model for individual home across several years and carefully study the impact of retrofits and renovations over time. Further, we can study the impact of any energy policy that the household participates.


\textbf{Homeowners:} Our approach  can provide custom recommendations to homeowners that best help reduce their energy footprint. When combined with geolocation data, homeowners can compare their efficiency to any region, including nearby neighborhoods. Such personalized energy reports can encourage consumers to take energy efficiency measures to reduce their footprint and energy costs. 


\section{Future Work}
\label{sec:futurework}

We now discuss \texttt{WattScale}'s strengths, limitations and future directions of our work. 
One of the strengths of WattScale is its applicability to large parts of the world. Since smart meters are being extensively deployed around the world, 
our approach can be used by tens of millions of homes that collect energy data. Further, our data-driven approach reduces the need for a full manual energy audit in homes. By identifying potential faults in homes, only a partial audit may suffice, thereby freeing resources and provide cost benefits.



In order for \texttt{WattScale} to provide accurate analysis, it requires energy consumption data at daily granularity. However, the building model can be modified to work with monthly energy bills, which are more widely available, especially in homes that do not have smart meters installed. Although the accuracy of the estimate of the building parameters may be lower in comparison to models built using daily energy. To overcome this limitation, one can use energy data across multiple years, which remains a part of our future work. 

The approach detailed in this work relies on comparing building parameters among similar homes. Currently, \texttt{WattScale} only looks at the following a building attributes --- (i) building age, (ii) size, and (iii) property type. In the New England dataset, we observed that building age and property type are proxies for several low-level features --- i.e., style of the building, flooring type, roof type, etc.  We believe that this is due to buildings adhering to the prevalent building codes of the time. However, one can also additionally use satellite data to augment our analysis. For example, we can learn if a home has a swimming pool that may require heating and a water pump, which increases its energy usage. Such homes can be compared to others with swimming pools for fair energy efficiency evaluation. We believe this is an interesting line of research and deserves more attention to gain new insights. Similarly, as part of future work, we intend also to utilize occupancy patterns a building attribute while creating the cohorts. For example, 24/7 occupancy homes should form a separate cohort.  

In the future, \texttt{WattScale} can also be enhanced to track energy savings and quantify the effectiveness of retrofits in homes. Moreover, 
while in our current work we look at residential buildings, our work can also be extended to identify inefficiencies in commercial buildings. Additionally, analyzing the seasonal changes (esp. weekly) could yield insights on energy usage patterns for different households. Such an analysis could provide feedback to the homeowners interested in knowing more about their energy consumption profile. For example, the energy data may reveal higher HVAC usage on Sundays, when homeowners are outdoors, thereby encouraging homeowners to set thermostats schedules.


%
%
%

\section{Related Work}
\label{sec:related}

Diagnosing and reducing energy consumption in buildings is an important problem~\cite{katipamula2005review, fontugne2013strip,zhou2009model, 
bellala2012following}.
Various methods have been proposed to detect abnormal energy consumption in a building~\cite{seem2007using,katipamula2005review, fan2014development}. 
However, these methods focused on commercial buildings that require expensive building management systems~\cite{seem2007using, fan2014development}
or requires costly instrumentation using sensors for monitoring purposes~\cite{bellala2012following, 
janetzko2014anomaly}. 
Sensors allow fine-grained monitoring of energy usage but are not scalable due to high installation costs. 
Unlike prior approaches, our model does not require building management systems or costly instrumentation and use ubiquitous smart meter data to determine energy inefficiency in buildings. 

Prior work have also proposed automatic modeling of residential loads~\cite{aftab2017real}. Studies have shown that compound loads  can be disaggregated into basic load patterns. Separately, there has been studies on non-intrusive load monitoring (NILM), which 
allow disaggregation of a household's total energy into its contributing appliances, and does not require building instrumentation~\cite{hart1992nonintrusive, 
batra2014comparison}. However, most NILM techniques require fine-grained datasets for training purposes and assume energy consumption patterns are similar across homes~\cite{batra2014comparison}. 
On the other hand, our approach makes no such assumption on energy consumption patterns and is applicable across multiple homes as it uses coarse-grained  energy usage
data that are readily available from utility companies~\cite{greenbutton}. 

Various energy performance assessment methods exist to quantify energy use in buildings and identify energy inefficiency~\cite{wang2012quantitative,yan2012simplified,hygh2012multivariate}. 
A common approach is to use degree-days method, a linear regression model, for calculating building energy consumption~\cite{
fels1986prism,
kissock2002development,
sambit}. 
However, these approaches do not consider  uncertainties that are associated with indicators of building performance. The idea of modeling uncertainties in thermal comfort is studied in \cite{
de1997influence}. But, it is
 restricted to a single office building with cooling and heating systems. 
Unlike previous studies, our approach can be used to identify least energy efficient home at scale without manual expert intervention.  
More recently, AI-based approaches have gained significant popularity in the energy and sustainability literature. Wang et al.~\cite{wang2017review} present a detailed review of AI-based models for energy usage in buildings. In our case, we propose a novel Bayesian model that has better interpretability as it accounts for uncertainties arising from human factors. 
Finally, we use actual ground truth data to validate our approach and show its efficacy on a large scale city-wide data.

\section{Conclusions}
\label{sec:conclusion}

Improving efficiency of buildings is an important problem, and the first step is to identify inefficient buildings. 
In this paper, we proposed \texttt{WattScale}, a data-drive approach to identify the least energy efficient homes in a city or region. 
We also implemented our approach as an open source tool, which we used to evaluate datasets from different geographical locations.  
We validated our approach on ground truth data and showed that our model correctly identified 95\% of the homes with inefficiencies. 
Our case study on a city-scale dataset using the individual execution mode showed that more than half of the buildings in our dataset are energy inefficient in one way or another,
of which almost 62.25\% of homes with heating related inefficiencies as probable cause. 
This shows that a lot of buildings can benefit from energy efficiency improvements. Further, as \texttt{WattScale} provided region-based execution mode that allows energy efficiency analysis of millions of homes in the US using publically available datasets.

As part of future work, we intend to deliver individual inefficiency report generated from our web application to the different homeowners. These nudges can be used to motivate and incentivize homeowners towards energy efficiency measures.

\begin{acks}
We thank all the reviewers for their insightful comments that helped us improve the paper. 
This research is supported by NSF grant CNS-1645952 and the Massachusetts Department of Energy Resources. 
 Any opinions, findings, and conclusions or recommendations expressed in this material are those of the author(s) and do not necessarily reflect
the views of the funding agencies. 
\end{acks}

\bibliographystyle{ACM-Reference-Format}
\bibliography{paper}

\end{document}